\documentclass[twocolumn,dvipsnames,twocolappendix]{aastex631}
\usepackage{amsmath, amsfonts, amssymb, graphicx, chngcntr, multirow, float}
\usepackage{mathrsfs}
\usepackage[]{hyperref}
\hypersetup{colorlinks=true}

\usepackage{booktabs}
\usepackage{amsmath}

\received{Feb. 28, 2025}
\revised{Mar. 19, 2025}
\accepted{Mar. 20, 2025}

\shorttitle{AT2022upj QPEs}
\shortauthors{Chakraborty \textit{et al.}}

\begin{document}

\title{Discovery of Quasi-periodic Eruptions in the Tidal Disruption Event and Extreme Coronal Line Emitter AT2022upj: implications for the QPE/TDE fraction and a connection to ECLEs}
\author[0000-0002-0568-6000]{Joheen Chakraborty}\thanks{joheen@mit.edu}
\affiliation{Department of Physics \& Kavli Institute for Astrophysics and Space Research, Massachusetts Institute of Technology, Cambridge, MA 02139, USA}

\author[0000-0003-0172-0854]{Erin Kara}
\affiliation{Department of Physics \& Kavli Institute for Astrophysics and Space Research, Massachusetts Institute of Technology, Cambridge, MA 02139, USA}

\author[0000-0003-4054-7978]{Riccardo Arcodia}\thanks{NASA Einstein Fellow}
\affiliation{Department of Physics \& Kavli Institute for Astrophysics and Space Research, Massachusetts Institute of Technology, Cambridge, MA 02139, USA}

\author[0000-0003-0426-6634]{Johannes Buchner}
\affiliation{Max Planck Institute for Extraterrestrial Physics, Giessenbachstrasse, Garching, 85748, Germany}

\author[0000-0002-1329-658X]{Margherita Giustini}
\affiliation{Centro de Astrobiolog\'ia (CAB), CSIC-INTA, Camino Bajo del Castillo s/n, 28692 Villanueva de la Ca\~nada, Madrid, Spain}

\author[0000-0002-8606-6961]{Lorena Hernández-García}
\affiliation{Millennium Nucleus on Transversal Research and Technology to Explore Supermassive Black Holes (TITANS)}
\affiliation{Millennium Institute of Astrophysics (MAS), Nuncio Monseñor Sótero Sanz 100, Providencia, Santiago, Chile}
\affiliation{Instituto de F\'isica y Astronom\'ia, Facultad de Ciencias, Universalized de Valpara\'iso, Gran Breta\~na 1111, Playa Ancha, Valpara\'iso, Chile}

\author[0000-0002-8304-1988]{Itai Linial}
\affiliation{Department of Physics and Columbia Astrophysics Laboratory, Columbia University, New York, NY 10027, USA}
\affiliation{Institute for Advanced Study, 1 Einstein Drive, Princeton, NJ 08540, USA}

\author[0000-0003-4127-0739]{Megan Masterson}
\affiliation{Department of Physics \& Kavli Institute for Astrophysics and Space Research, Massachusetts Institute of Technology, Cambridge, MA 02139, USA}

\author[0000-0003-0707-4531]{Giovanni Miniutti}
\affiliation{Centro de Astrobiolog\'ia (CAB), CSIC-INTA, Camino Bajo del Castillo s/n, 28692 Villanueva de la Ca\~nada, Madrid, Spain}

\author{Andrew Mummery}
\affiliation{Oxford Theoretical Physics, Beecroft Building, Clarendon Laboratory, Parks Road, Oxford, OX1 3PU, United Kingdom}

\author[0009-0001-9034-6261]{Christos Panagiotou}
\affiliation{Department of Physics \& Kavli Institute for Astrophysics and Space Research, Massachusetts Institute of Technology, Cambridge, MA 02139, USA}

\author[0000-0002-7116-2897]{Erwan Quintin}
\affiliation{European Space Agency (ESA), European Space Astronomy Centre (ESAC), Camino Bajo del Castillo s/n, 28692 Villanueva de la Cañada, Madrid, Spain}

\author[0000-0003-0820-4692]{Paula Sánchez-Sáez}
\affiliation{European Southern Observatory, Karl-Schwarzschild-Strasse 2, 85748 Garching bei München, Germany}

\begin{abstract}
Quasi-periodic eruptions (QPEs) are recurring soft X-ray transients emerging from the vicinity of supermassive black holes (SMBHs) in nearby, low-mass galaxy nuclei; about ten QPE hosts have been identified thus far. Here we report the \textit{NICER} discovery of QPEs in the optically-selected Tidal Disruption Event (TDE) and Extreme Coronal Line Emitter (ECLE) AT2022upj, exhibiting a large spread in recurrence times from 0.5-3.5 days, durations from 0.3-1 days, peak luminosities from $10^{42.5-43.0}$ erg s$^{-1}$, and erratic flare profiles. A wealth of evidence now links at least some QPEs to the newly-formed accretion flows emerging from TDEs; AT2022upj is the third QPE reported in an optically-discovered TDE. Marginalizing over the uncertain distributions of QPE peak luminosity, recurrence time, delay after TDE peak, and lifetime, we use the burgeoning sample to make a Bayesian estimate that the fraction of optical TDEs resulting in QPEs within 5 years post-disruption is $9^{+9}_{-5}$\%. Along with AT2019qiz, AT2022upj also marks the second of the three optical TDE+X-ray QPEs showing coronal line emission, suggesting ECLEs may represent a subset of TDEs particularly efficient at forming QPEs and/or that sustained QPE X-ray emission contributes to coronal line emission in some galaxy nuclei.
\end{abstract}


\section{Introduction} \label{sec:intro}
Quasi-periodic eruptions (QPEs) are recurring soft X-ray transients from supermassive black holes (SMBHs), observed from about ten galaxies thus far \citep{Miniutti19,Giustini20,Arcodia21,Chakraborty21,Quintin23,Arcodia24a,Nicholl24,Hernandez25}. They show peak luminosities ranging from $L_{\rm peak}\sim 10^{42-44}$ erg s$^{-1}$, recurrence times of $T_{\rm rec}\sim 2.5-100\;$hr, blackbody-like spectra with temperatures of $kT\sim 50-250\,$ eV, SMBH masses of $\sim 10^{5-7.5}M_\odot$, and host galaxy redshifts up to $z\sim 0.08$.

In the 5 years following the discovery of QPEs, several observational coincidences linked them to the aftermath of TDEs. The discovery of QPEs in the 2019 occurred in the galaxy GSN 069, which showed a sudden $\geq 240\times$ X-ray brightening in 2010 followed by a decade-long decline \citep{Miniutti13,Miniutti19}; the same source has since exhibited \textit{another} long-term brightening and decline, suggesting a potential repeating TDE \citep{Miniutti23a}. Following the discovery in GSN 069, QPEs were blindly detected in another source from the \textit{XMM-Newton} archives, which was also a TDE candidate showing a $\sim 100\times$ brightening and long-term decline \citep{Chakraborty21}. It is also noteworthy that several (but not all) known QPEs show a secular decreases in their quiescence and peak X-ray luminosities \citep{Arcodia24a,Chakraborty24,Pasham24a,Giustini24}.

Some QPEs show evidence for abnormally high C/N ratio in their ultraviolet \citep{Sheng21} and X-ray spectra \citep{Kosec25,Chakraborty25}; a similar enhancement is also seen in some TDEs, perhaps due to CNO processing in the disrupted stellar envelope \citep{Cenko16,Mockler22,Miller23}. GSN 069 and ZTF19acnskyy/Ansky host compact nuclear [O III]-emitting regions of $<$35 pc and 1.1 pc respectively, suggesting young accretion systems $<100$ years old \citep{Patra24,Sanchez24} and nonstellar nuclear ionization regions \citep{Wevers24}. Moreover, none of the QPE host galaxies show broad optical/UV emission lines despite a lack of intrinsic obscuration in those sources \citep{Wevers22}, which is incompatible with standard photoionization processes from extended AGN accretion disks supporting broad line regions. UV to X-ray SED modeling of multiple QPE hosts has also revealed their accretion disks to be compact---$\mathcal{O}(10^{2-3}R_g)$ in radial extent---with evidence of significant viscous spreading on year timescales, as expected for TDE disks \citep{Mummery20,Nicholl24,Guolo25,Wevers25}. QPEs and TDEs are significantly overrepresented in post-starburst and quiescent Balmer-strong galaxies, which comprise $\lesssim0.2\%$/$2\%$ of galaxies in the local universe respectively, but comprise a third of known QPEs \citep{Wevers24} and $\sim 32$\% of TDEs \citep{French20}. Studies of the host galaxy morphologies of QPEs and TDEs find both populations exhibit anomalously high Sérsic indices, $B/T$ ratios, and surface mass densities \citep{Gilbert24}. These extreme statistical coincidences among both populations suggest a common link to the gas-rich, centrally concentrated environments of recently faded galactic nuclei.

There is now direct evidence that \textit{at least some} QPEs emerge in bona-fide TDEs. Likely QPEs were reported in AT2019vcb \citep{Quintin23,Bykov25}, with the first definite detection in AT2019qiz \citep{Nicholl24}; both are optically-discovered, spectroscopically-confirmed TDEs from the Zwicky Transient Facility (ZTF) survey \citep{Bellm19}. In the years after optical peak, both optical TDEs showed X-ray QPEs as the overall optical/UV flux declined. The discovery of QPEs in Ansky \citep{Hernandez25} following a potential AGN turn-on (or another class of unknown SMBH transient) suggests that a broader class of newborn accretion flows may support QPEs, and that the TDEs are not the \textit{only} way to attain the necessary conditions.

There are compelling theoretical ground for some QPE models to favor TDEs, rather than long-lived AGN, for the underlying disk. The current leading theoretical paradigm invokes a stellar-mass companion captured by the gravitational influence of an SMBH in an extreme mass-ratio inspiral (EMRI; \citealt{Linial23a,Franchini23,Xian21,Linial23b,Tagawa23,Zhou24a}); over the duration of its orbit, the companion collides with the SMBH accretion flow, leading to a high-temperature shock breakout whose evolution produces significant X-ray emission \citep{Vurm24,Chakraborty25}. These models elegantly explain many of the observational properties of QPEs, such as the stable long-/short-alternation between recurrence times seen in some QPEs \citep{Miniutti13,Arcodia24c,Pasham24b} and the spectral evolution which shows a strikingly similar color-dependence to shock-powered supernovae \citep{Kasen06}. The measured QPE blackbody emission areas may also imply the emission is initially generated from $R_{\mathrm{bb}}\sim R_\odot$, compatible with originating from roughly a stellar cross-section (though these measurements can deviate by factors of several from any true physical radius due to scattering-dominated atmospheres and non-LTE effects; \citealt{Linial23b}).

One prediction of stellar EMRI models is gradual ablation of the companion by repeated disk collisions over $\mathcal{O}$(years-decades) \citep{Linial24b,Yao25}. This naturally explains why QPEs prefer TDE disks, rather than AGN which are more intrinsically common: the long-lived, radially extended disks of AGN would destroy EMRIs at orbital periods $\gg\:$hrs-days ($R \gg 100-500 R_g$), long before the compact orbits required for QPEs. Thus QPEs would require a ``dry'' EMRI unimpeded by environmental effects (e.g. an extended accretion disk), until the companion reaches 100-500 $R_g$. We emphasize that this is only one QPE model paradigm: not all theories necessitate a QPE-TDE association, so it is observationally important to constrain this occurence rate and distinguish between models. There are other flavors of EMRI models \citep{King22,Krolik22,Lu23,Yang25} as well as disk-instability models which need not invoke any orbiter altogether \citep{Raj21,Pan22,Middleton25}, and further tests confronting observational data against the predictions of these models are necessary to confirm the true underlying driver of QPEs. 

Here we report the discovery of QPEs in a third optically discovered TDE, AT2022upj. AT2022upj was first reported as an optical transient coincident with the nucleus of a galaxy at a redshift $z=0.054$ by ZTF on August 31, 2022. Estimates of the black hole mass from TDE light curve fits and host-galaxy scaling relations range between $M_\bullet\approx(0.6-1.6)\times 10^6M_\odot$ depending on the method used \citep{Newsome24}. Follow-up optical spectroscopy revealed broad, rapidly evolving hydrogen Balmer lines, a distinguishing signature seen in the majority of TDEs.  \textit{Unlike} most TDEs however, AT2022upj showed the prompt emergence of bright coronal emission lines---with ionization states of [Fe\,{\sc vii}]-[Fe\,{\sc xi}]---indicating higher-than-typial circumnuclear gas content ($10^7-10^{10}$ cm$^{-3}$) ionized by the newborn accretion flow for the transitions to occur at detectable rates. While such ``extreme coronal line emitters'' (ECLEs) have been linked suggestively with  TDEs for over a decade \citep{Wang12} due to statistical trends in their host galaxy properties and the presence of secular declines in their line fluxes, AT2022upj joins the small population of ECLEs directly associated with known TDEs. It is also the first ECLE+TDE to show immediate coronal lines accompanying the disruption.

In Section~\ref{sec:results} we present the light curves and spectra indicating the discovery of QPEs in AT2022upj, along with accretion disk fits of the late-time plateau to assess whether the disk radial size is large enough to account for QPEs in the orbiter-disk collision picture. In Section~\ref{sec:qpe_tde_rate} we discuss the sample of optical TDEs followed by X-ray QPEs, now three members large, and make a preliminary estimate of the occurrence fraction of QPEs following optical TDEs. In Section~\ref{sec:discussion} we compare to independently derived QPE volumetric rates from prior work, and remark on the emerging coincidence of QPEs in ECLEs. In Section~\ref{sec:conclusion} we make concluding remarks and highlight future prospects for follow-up study. Details of the observations and data reduction procedures can be found in Appendix~\ref{sec:methods}.

\section{Results} \label{sec:results}

\begin{figure*}
    \centering
    \includegraphics[width=\textwidth]{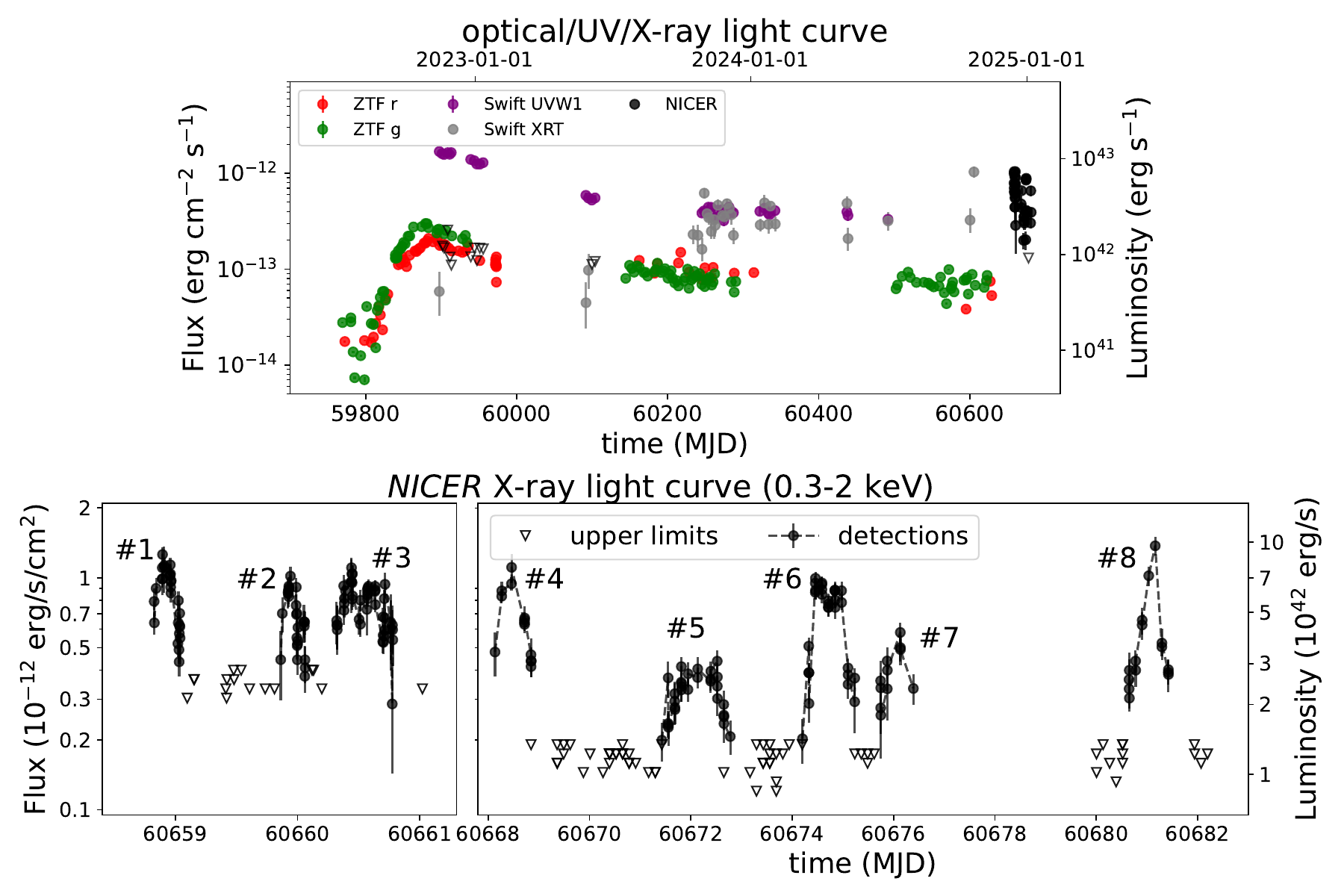}
    \caption{\textbf{Top:} Long-term optical (ZTF), UV (\textit{Swift} UVOT), and X-ray (\textit{Swift}, \textit{NICER}) light curves of AT2022upj. Although the X-ray emission at early times was faint, the source showed a delayed X-ray brightening $\sim 1$ year post-optical/UV peak. The first indication of QPEs occurred another $\sim$year after. \textbf{Bottom:} \textit{NICER} X-ray light curve from Dec. 12, 2024-Jan. 8, 2025 revealing quasi-periodic eruptions with 0.5-3.5 day recurrence times.}
    \label{fig:lc}
\end{figure*}

In the top panel of Fig.~\ref{fig:lc} we show the long-term optical (ZTF g-/r-band), UV (\textit{Swift} UVW1), and X-ray (0.3-10 keV) light curves of AT2022upj. The optical/UV TDE peak occurred in late 2022, at which time only faint X-ray emission was detected. The optical/UV emission then declined before reaching a plateau at a higher luminosity than the pre-outburst level, indicating the formation of a compact nuclear accretion disk \citep{vanVelzen19}. Following this, in late 2023 the X-ray flux rose by about $3-4\times$, and began exhibiting days-timescale variability with a characteristic amplitude of $\sim 2\times$. These data were already reported and analyzed in \cite{Newsome24}, which also studied the unusually prompt appearance of coronal lines in the optical spectra.

In the bottom panels of Fig.~\ref{fig:lc}, we zoom into the recent months of \textit{NICER} X-ray data, and show the light curve generated by our time-resolved spectroscopy approach outlined in Sec.~\ref{subsec:nicer}. In total, eight QPEs were observed (labeled in the figure), with characteristic peak amplitudes of $\geq 5-10\times$ quiescence, a large spread in recurrence times ranging from $\sim 0.5-3.5$ days, and durations $\sim 0.3-1$ day. The current quiescent level is not detected by \textit{NICER}. The flare profiles do not always show an exponential rise-and-decline shape as is seen in most other QPEs; they are comparatively erratic, with potentially some overlapping bursts (e.g. QPEs \#3 and \#6 in Fig.~\ref{fig:lc}) akin to the behavior observed in eRO-QPE1 \citep{Arcodia22,Chakraborty24}. The large spread in recurrence times also mirrors the irregularity exhibited by RXJ1301 \citep{Giustini24}.

\begin{figure}
    \centering
    \includegraphics[width=\linewidth]{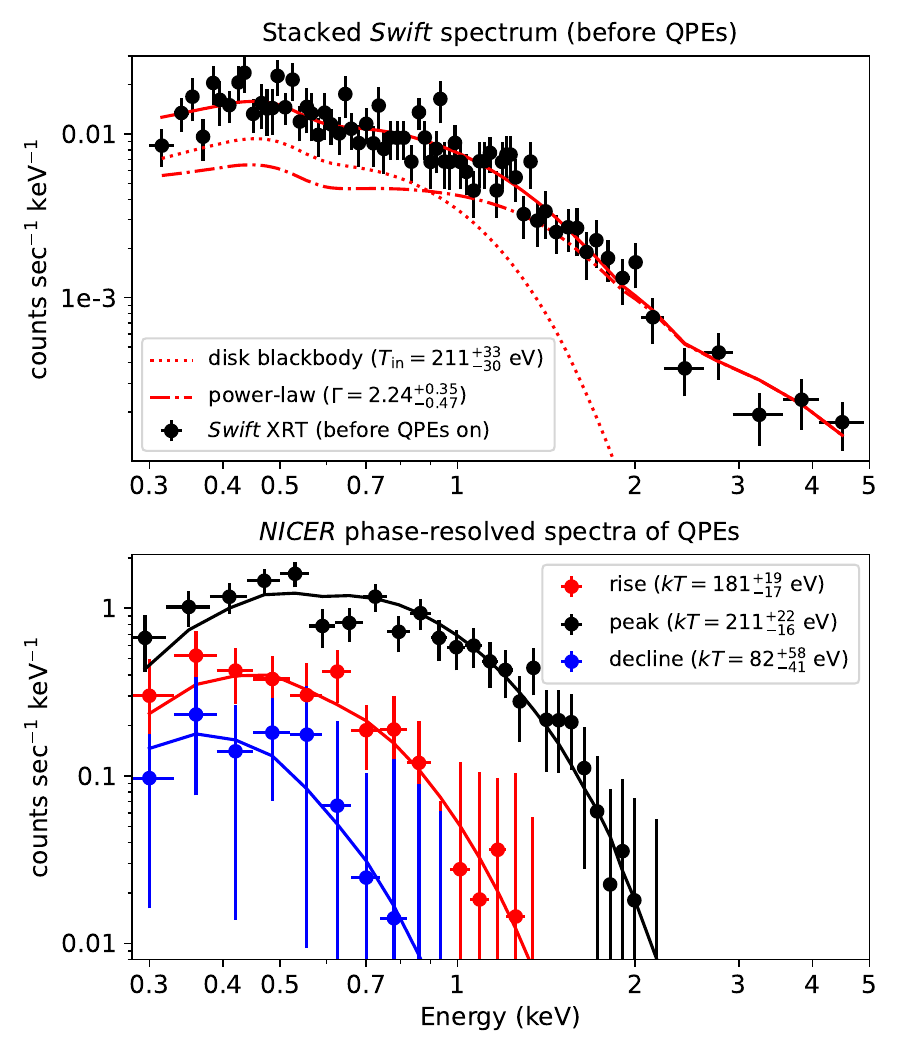}
    \caption{\textbf{Top:} Stacked \textit{Swift} XRT spectrum (72770~ks) of the late-time rise (between MJDs 60200-60600 in Fig.~\ref{fig:lc}). We show fits with a disk blackbody + power-law model (Table~\ref{tab:spec}). \textbf{Bottom:} Background-subtracted \textit{NICER} spectra of QPE \#4 (Fig.~\ref{fig:lc}) fit with a blackbody model for the rise, peak, and decline phases (Table~\ref{tab:spec}).}
    \label{fig:spec}
\end{figure}

In the top panel of Fig.~\ref{fig:spec} we show the stacked \textit{Swift} XRT spectrum of the late-time slowly-evolving phase between MJDs 60200-60600, before the QPEs were detected. The delayed emergence of X-rays is a common phenomenon seen in TDEs \citep{Guolo24b} and is likely associated with a long viscous timescale in the accretion disk. The X-ray emission shows a significant hard power-law component of comparable flux to the thermal blackbody as seen in a handful of TDEs \citep{Yao22,Guolo24b}, but with a significantly higher luminosity than in other QPEs. It is unclear why AT2022upj shows such a prominent power-law component compared to other QPEs, and it is an interesting avenue for follow-up study. We note that our \textit{NICER} data are not able to constrain whether the power-law is still present in the QPE spectra, as the observational background dominates for energies $\gtrsim 1$ keV at the current flux level of AT2022upj. We fit the stacked \textit{Swift} spectrum with a composite disk blackbody and hard X-ray power-law model, accounting for host galaxy redshift ($z=0.054$) and ISM absorption, i.e. a total model of \texttt{tbabs}$\times$\texttt{zashift}$\times($\texttt{diskbb}$+$\texttt{powerlaw}$)$; the resulting parameters are reported in Table~\ref{tab:spec}. We note that, given the significant hard X-ray flux and high inner disk temperature of the thermal component, we would likely \textit{not be sensitive} to the presence of QPEs promptly after the delayed X-ray rise. It is thus possible that we are only seeing QPEs emerge now as the quiescence X-ray luminosity has declined sufficiently for the X-ray bursts to stand out.

The QPE spectra themselves are soft and thermal-like, as is the case in all other sources. The majority of the bursts are not  suitable for spectral fitting for one or multiple reasons: eruptions \#1-3 were observed during \textit{NICER} orbit day, meaning the effects of optical loading significantly limit the usable low-energy cutoff and increase uncertainty in the instrument response; bursts \#5 and 7 were lower flux, reaching the limit of signal-to-noise for which spectral fit parameters can be reliably estimated; and burst \#8 is poorly-sampled, with a lower effective cadence of good-quality exposures on the source. We thus limited detailed spectral study to bursts \#4 and 6; the resulting fits are presented in Table~\ref{tab:spec}. In Fig.~\ref{fig:spec} we show phase-resolved \textit{NICER} spectra during the rise, peak, and decline of QPE \#4, which is comparatively well-behaved compared to the other burst profiles. The rise/peak/decline blackbody temperatures are $kT=181/211/82$ eV, respectively, which is relatively high-temperature compared to the usual peaks around 100-150 eV, but not unprecedented as some QPEs exhibit temperatures up to 250 eV \citep{Arcodia21}.

\begin{figure}
    \centering
    \includegraphics[width=\linewidth]{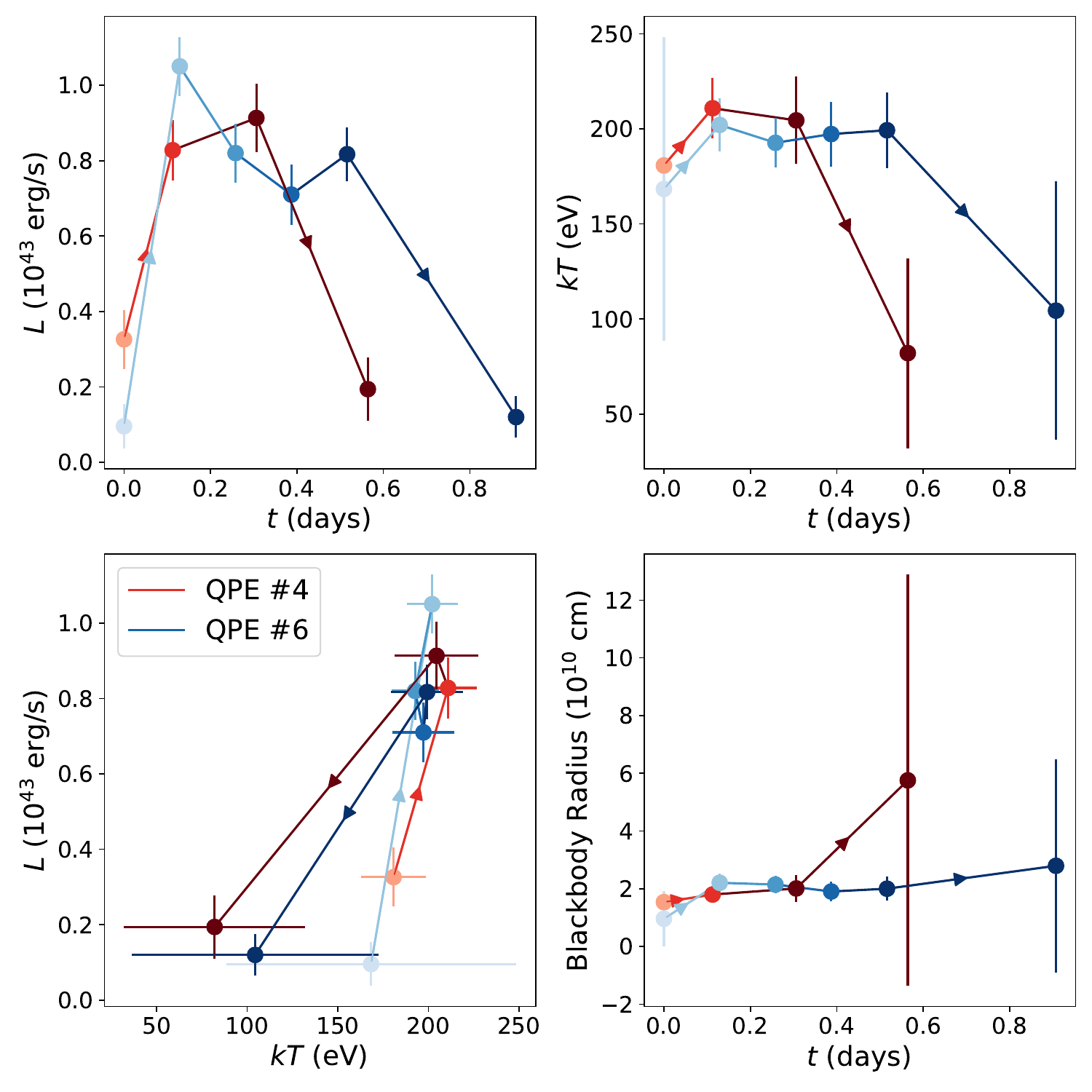}
    \caption{Spectral evolution of the QPEs in AT2022upj for bursts \#4 (red) and 6 (blue). The latter eruption shows a more complex eruption profile reminiscent of the overlapping bursts of eRO-QPE1, whereas burst \#4 shows the more standard hysteresis behavior seen in QPEs (bottom left panel).}
    \label{fig:spec_evol}
\end{figure}

\begin{table*}
\centering
\caption{Spectral Fitting Results. \textbf{Top:} Physical parameters from the stacked \textit{Swift}-XRT spectrum before QPEs turned on (MJDs 60200--60600). We used a total model of \texttt{tbabs}$\times$\texttt{zashift}$\times($\texttt{diskbb}$+$\texttt{powerlaw}$)$ with $N_{\rm H}=2.13\times 10^{20}$ cm$^{-2}$ and $z=0.054$. \textbf{Bottom:} \textit{NICER} phase-resolved spectral fitting results for the QPEs on MJD 60668 and 60674 (see Fig.~\ref{fig:lc}). We used a model of \texttt{tbabs}$\times$\texttt{zashift}$\times$\texttt{bbody}.}
\label{tab:spec}
\begin{tabular}{lccccc}
\toprule
\multicolumn{6}{c}{Stacked \textit{Swift} XRT spectrum before QPEs} \\
\midrule
 & $T_{\rm in}$ & $L_{\rm disk,bol}$ & $\Gamma$ & $L_{\rm pow,bol}$ & C-stat/dof \\
 & (eV) & ($10^{42}$ erg s$^{-1}$) & --- & ($10^{42}$ erg s$^{-1}$) & --- \\
\midrule
 \multirow{2}{*}{\shortstack{Quiescence (before QPEs) \\ \texttt{tbabs}$\times$\texttt{zashift}$\times($\texttt{diskbb}$+$\texttt{powerlaw}$)$}} & $211^{+33}_{-30}$ & $2.65^{+0.31}_{-0.30}$ & $2.24^{+0.35}_{-0.47}$ & $7.43^{+0.72}_{-0.67}$ & 177/213 \\
 & & & & & \\
\midrule
\midrule
\multicolumn{6}{c}{\textit{NICER} phase-resolved spectral fits during QPEs} \\
\midrule
 & Time & $kT$ & $L_{\rm bol}$ & $R_{\rm bb}$ & C-stat/dof \\
 & (MJD) & (eV) & ($10^{42}$ erg s$^{-1}$) & ($10^{10}$ cm) & --- \\
\midrule
\multirow{4}{*}{\shortstack{QPE \#4 \\ \texttt{tbabs}$\times$\texttt{zashift}$\times$\texttt{bbody}}} & 60668.15 & $181^{+19}_{-17}$ & $3.26^{+0.79}_{-0.77}$ & $1.54^{+0.36}_{-0.34}$ & 73/84 \\ 
 & 60668.26 & $211^{+22}_{-16}$ & $8.28^{+0.82}_{-0.74}$ & $1.80^{+0.28}_{-0.29}$ & 87/80 \\
 & 60668.46 & $205^{+19}_{-16}$ & $9.13^{+0.70}_{-0.70}$ & $2.01^{+0.48}_{-0.46}$ & 79/84 \\
 & 60668.71 & $82.1^{+58}_{-41}$ & $1.94^{+0.85}_{-0.84}$ & $5.76^{+7.2}_{-7.1}$ & 59/77 \\
 \midrule
\multirow{6}{*}{\shortstack{QPE \#6 \\ \texttt{tbabs}$\times$\texttt{zashift}$\times$\texttt{bbody}}} & 60674.58 & $168^{+84}_{-80}$ & $5.81^{+0.60}_{-0.51}$ & $0.959^{+0.96}_{-0.95}$ & 85/74 \\
 & 60674.71 & $202^{+19}_{-13}$ & $10.5^{+0.89}_{-0.82}$ & $2.21^{+0.32}_{-0.30}$ & 128/79 \\
 & 60674.84 & $193^{+18}_{-15}$ & $8.20^{+0.79}_{-0.74}$ & $2.14^{+0.30}_{-0.31}$ & 117/79 \\
 & 60674.45 & $197^{+23}_{-22}$ & $7.13^{+0.79}_{-0.76}$ & $1.90^{+0.35}_{-0.34}$ & 96/81 \\
 & 60674.33 & $199^{+21}_{-12}$ & $8.17^{+0.73}_{-0.71}$ & $2.00^{+0.42}_{-0.42}$ & 104/81 \\
 & 60675.23 & $104^{+78}_{-63}$ & $1.20^{+0.62}_{-0.51}$ & $2.80^{+3.8}_{-3.7}$ & 102/77 \\
\bottomrule
\end{tabular}
\end{table*}

In Fig.~\ref{fig:spec_evol}, we show the co-evolution of QPE spectral parameters for bursts \#4 and 6. In the top-left panel, we show the bolometric luminosities over time, which were computed from 0.001-100 keV using the \texttt{clumin} model in \texttt{XSPEC}. In the top-right,  we show blackbody temperatures ($kT$) in eV. In the bottom-left, we show bolometric luminosity against temperature, indicating the ``hysteresis'' behavior seen ubiquitously in QPEs, whereby the rise is higher-temperature than the decline. This spectral evolution is typically considered an identifying signature of QPEs, as it is seen in all known sources despite variations in other observables, indicating a common physical process driving their emission. In the bottom-right panel, we show the inferred blackbody radius from the Stefan-Boltzmann law, $R_{\rm bb} = \sqrt{L_{\rm bol}/4\pi\sigma_{\rm SB}T^4}$. We report the resulting spectral fit parameters in Table~\ref{tab:spec}. The evolution in the overlapping rises of burst \#6 show a qualitatively similar evolution to eRO-QPE1 \citep{Arcodia22}: the temperature increases most dramatically during the first rise phase, then again slightly in the second. The inferred blackbody emission radii are of order $R_{\rm bb} \sim R_\odot$, as is also generally the case in QPEs, with a slight increasing trend from the beginning to end of the burst. The measured blackbody radii need not directly correspond to any true physical scale within the system due to scattering-dominated atmospheres and non-LTE effects \citep{Linial23b}, but the increasing trend in $R_{\rm bb}$ is seen ubiquitously in QPEs. Within the collisional model, this is attributed to expansion of the emission surface, which is a basic expectation in the case that QPE emission is powered by shock breakouts following orbiter-disk \citep{Vurm24}.

\begin{figure}
    \centering
    \includegraphics[width=\linewidth]{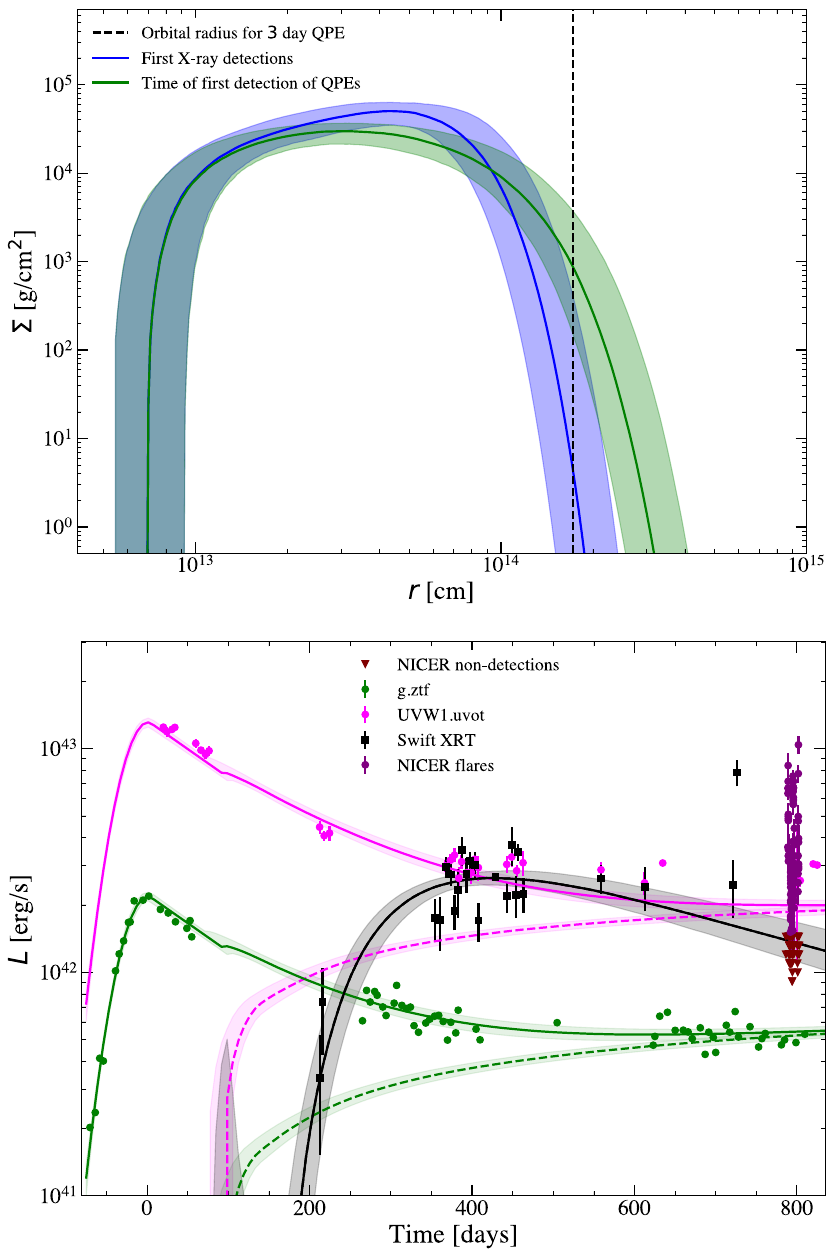}
    \caption{\textbf{Top:} inferred accretion disk surface density profiles $\Sigma(r)$ at the times of the delayed X-ray rise (blue) and the emergence of QPEs (green). The inferred outer radius is consistent with viscously expanding over $\sim$few hundred days to the orbital radius corresponding to a 3-day QPE. \textbf{Bottom:} model light curves fit to the \textit{Swift} UVW1, ZTF g-band, and \textit{Swft} X-ray data. The presence of the optical/UV plateaus and delayed X-ray rise indicate the delayed formation, and viscous expansion, of an accretion disk.}
    \label{fig:disk_model}
\end{figure}

A striking feature of the multi-band light curve is the delayed optical/UV plateau at a luminosity higher than the pre-outburst state, which suggests the formation of a compact accretion disk as seen in at least two-thirds of TDEs \citep{vanVelzen19,Mummery20, Mummery24a}. At times $t \gtrsim 60200$ MJD all of the observed emission from the source is dominated by direct accretion disk emission, offering an opportunity to constrain the properties of the disk during the time when QPEs were detected. We used the time-dependent general relativistic thin disk {\tt FitTeD} code of \cite{Mummery24b} to fit the binned ZTF g-band, \textit{Swift} UVW1-band, and \textit{Swift} X-ray data to estimate the properties of this late-time disk. The same analysis was performed for the late-time accretion disk of AT2019qiz in \cite{Nicholl24}. A corner plot of the fitted parameters are displayed Appendix~\ref{subsec:corner_plots}.

In the top panel of Fig.~\ref{fig:disk_model}, we show the radial surface density profile at the time of the delayed X-ray rise (blue), then once the QPEs began (green). Suggestively, the best-fit model is consistent with a disk containing $\sim 0.5M_\odot$ of material viscously expanding over the $\sim 200$ days between the X-ray rise and the first indication of QPEs to reach approximately the predicted orbital radius.  While it is a model-dependent claim, this tentatively aligns with the predictions of the orbiter-disk collision scenario, as QPEs can only emerge once the accretion disk is radially extended enough to interact with the EMRI. In the bottom panel of Fig.~\ref{fig:disk_model} we show the output model luminosity alongside the optical/UV/X-ray data point used for the fits. The early time light-curve is dominated by the prompt emission, which is generated by a uncertain processes \citep[see e.g.,][for a discussion of the uncertainties in early optical emission in TDEs]{Roth20_ISSI}. Existing models include reprocessing or collisional shocks from self-intersecting debris streams, but it is known that this emission does not result directly from an accretion flow; we  parametrize our ignorance of this initial phase with an exponential decay. The dashed lines indicate the luminosity contributed by the accretion disk, which is negligible at early times but eventually becomes the dominant source driving the late-time plateau. The \textit{NICER} upper limits after the QPEs began are consistent with the quiescence luminosity predicted by the disk viscous expansion. We note that the X-ray emission from the disk model is strictly thermal, even though we know there is a comparable X-ray flux in the power-law as seen in the top panel of Fig.~\ref{fig:spec}. However, this factor of $\sim 2\times$ difference will not make any significant change to the inferred disk radius or mass. This is because in the Wien tail the X-ray flux is exponentially sensitive to the disk mass \citep[meaning that changes in the inferred disk mass will go only logarithmically with changes in the flux][]{Mummery21}, while the disk outer radius is constrained entirely by the optical/UV flux.  We show the full corner plot of the disk parameters resulting from our light curve fits in Appendix~\ref{subsec:corner_plots}. Our fit also produces an independent SMBH mass estimate of $M_\bullet \sim 5.6^{+3.3}_{-2.5}\times 10^6M_\odot$. The various mass estimates of \cite{Newsome24} reported a range of $(0.6-1.6)\times10^6\;M_\odot$; while our value is higher than this, $M_{\rm BH}$ scaling relations are typically accompanied by large scatter up to $\sim 1$ dex, making the tension statistically insignificant.

\begin{figure}
    \centering
    \includegraphics[width=\linewidth]{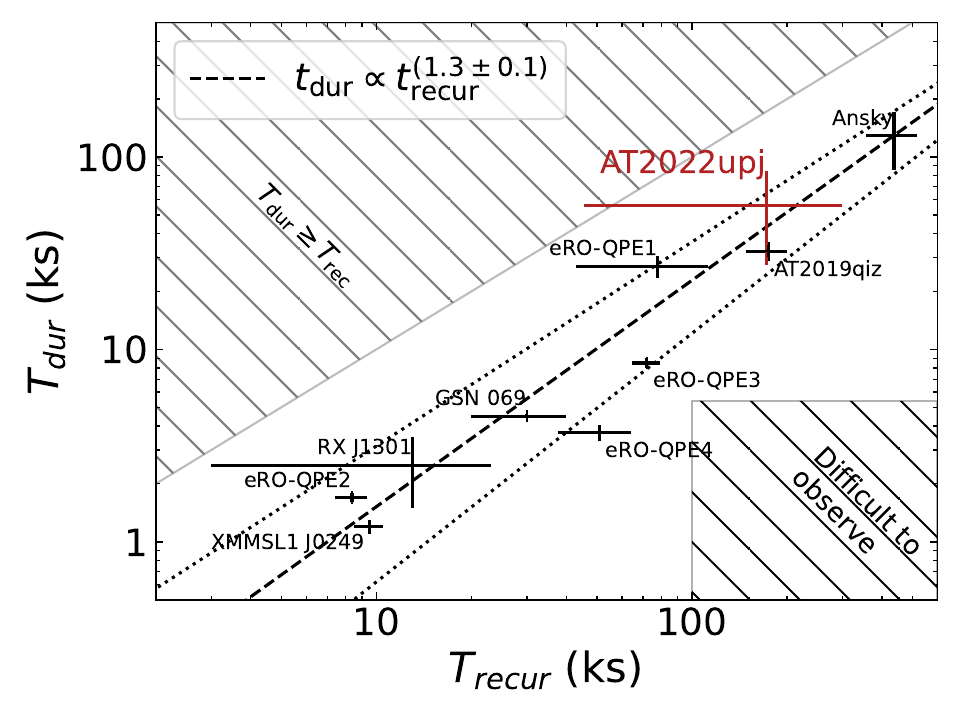}
    \caption{Correlation of QPE burst duration ($T_{\rm dur}$) with recurrence time ($T_{\rm rec}$). The error bars show the diversity of timescales in each source (not measurement uncertainties). AT2022upj shows a relatively high scatter in both quantities. The top-left hatched region denotes $T_{\rm dur}\geq T_{\rm rec}$, which is an invalid region of parameter space. The bottom-right region denotes $T_{\rm dur}<16$~ks and $T_{\rm rec}>100$~ks, which is rendered unobservable by current instrument capabilities.}
    \label{fig:tdur_trecur}
\end{figure}

In Fig.~\ref{fig:tdur_trecur}, we show the QPE duration ($T_{\rm dur}\sim 0.3-1$ days) and recurrence time ($T_{\rm rec}\sim 0.5-3.5$ days) of AT2022upj in comparison with other known QPEs. We also shade excluded regions of parameter space. For instance, sources having $T_{\rm dur}\geq T_{\rm rec}$ would not be identifiable as QPEs, as they would not have any well-resolved individual bursts. Moreover, the current observational capabilities of \textit{XMM-Newton} and \textit{NICER}---which have thus far been responsible for the confirmation of all known QPEs---face a coverage gap for roughly $T_{\rm dur}\lesssim 5.4$~ks and $T_{\rm rec}\gtrsim 100$~ks. Short-duration sources can only be resolved by the continuous monitoring capabilities of \textit{XMM-Newton}, due to the 90-minute ISS orbit inducing periodic gaps in \textit{NICER} visibility. On the other hand, long-recurrence sources can only be resolved by \textit{NICER}, as the elliptical 48-hour orbit of \textit{XMM-Newton} limits its ability to continuously monitor a source for $\gtrsim 100$~ks. Sources falling in both ``blind spots'', with durations too short for \textit{NICER} and recurrence times too long for \textit{XMM-Newton}, would likely be missed by current observational capabilities. The overall population roughly follows a scaling relation of $T_{\rm dur}\propto T_{\rm rec}^{1.3}$, but we caution against over-interpreting this apparent correlation as 1) it is still subject to small-number statistics, and 2) it may be driven in part by observational coverage gaps rather than any true physical link. 

Despite its erratic behavior and complex eruption profiles, AT2022upj appears to align with this relationship, with longer-than-average $T_{\rm dur}$ and $T_{\rm rec}$ compared to detected QPEs. From the X-ray light curve of Fig.~\ref{fig:lc}, it is not clear that there is a relationship between the quiescence before/after a burst and its duration. In some instances, there is a long pause before a short eruption (e.g. QPEs \#2-3), while in others there is a short pause followed by a short eruption (e.g. QPEs \#6-7). There is also not a clear relationship between the burst peak luminosity and timescale, as QPE \#5 is both particularly long-duration and low-luminosity. The bursts of eRO-QPE1 and RX J1301 can be similarly unpredictable, with no hard-and-fast correlations between duration, separation, and luminosity \citep{Giustini24,Chakraborty24}. The irregularity of AT2022upj also bears similarities to the longer-timescale bursts of Swift J023017.0+283603 \citep{Evans23b,Guolo24a}, which show variable flare profiles and ``missed'' eruptions. On the other hand, Swift J0230 does not fall exactly on the $T_{\rm dur}-T_{\rm rec}$ trend of Fig.~\ref{fig:tdur_trecur}---its recurrence time is several times longer than would be predicted by its duration---and it does not show the same spectral hysteresis characterizing the QPEs, indicating a different emission mechanism. AT2022upj may represent a subclass of sources with intermediate properties, with similar radiation processes to standard QPEs, but with some properties (e.g. orbital eccentricity, disk profile) similar to repeating TDEs.

There are a handful of possible explanations for the irregular QPE timings in AT2022upj within the orbiter-disk collision paradigm, which at first sight, should produce relatively stable clocks set by the orbital period. Faced with the same tension in eRO-QPE1, \cite{Chakraborty24} applied the \textit{O-C} technique to analyze the residuals from a fixed underlying period, finding super-periodic, sinusoidal-like structure with a period of $\sim 6$ days. This can be reasonably attributed to Lense-Thirring precession of the underlying accretion disk \citep{Franchini16} for a degenerate (but plausible) range of density profiles and SMBH spins. In eRO-QPE1, the spread in QPE timings can be as high as 50\% the recurrence time, which is similar to the range exhibited by AT2022upj. This amplitude, in the rigidly-precessing accretion disk picture, would be set primarily by the disk inclination, and may be expected to decrease with time as it comes into alignment with the SMBH spin. The \textit{O-C} analysis has also been applied to GSN 069 in \cite{Miniutti25}, which also found super-periodic structure with a smaller amplitude of $\sim20\%$ the recurrence time. Though the sample of QPEs to which this analysis has been applied is small, it is noteworthy that both cases found similar results, and quasi-sinusoidal modulation of the burst timings with a range of different amplitudes may be a common phenomenon in QPEs. Continued monitoring of the bursts in AT2022upj, ideally with longer baselines, will directly test whether any structure can be found in the apparently chaotic bursts of AT2022upj, as has been done for the others.

Another feasible explanation was put forth in \cite{Yao25}, which suggested that some systems may be dominated by \textit{debris}-disk interactions rather than star-disk collisions. In those cases, the spread in debris energy would result in a dispersion of the collision times, which can easily produce a wide range of phenomenology including overlapping bursts and timing deviations of 50\% for sufficiently large orbits (see their Eq. 8). Other EMRI models can also produce 
deviations from strict periodicity. For instance, white dwarf mass-transfer models experience significant angular momentum losses in the accreted material, which can appreciably modify the orbital energies and radii, whereas relativistic effects near pericenter---particularly for the high-eccentricity EMRIs invoked for these models---can produce significant precession of just a few orbits \citep{Yang25}. Disk instability models offer perhaps the greatest flexibility in this regard, as uncertainties in the accretion process in general, let alone the exotic subset which may be responsible for QPEs, are large enough that a wide range of timescales can be accommodated by varying the local viscosity and heating/cooling rate \citep{Pan22}.

\section{The fraction of optical TDEs resulting in QPEs} \label{sec:qpe_tde_rate}
AT2022upj is the third optically selected TDE showing QPEs, following AT2019vcb \citep{Quintin23,Bykov25} and AT2019qiz \citep{Nicholl24}. Late-time follow-up of X-ray and optically-selected TDEs is therefore a promising avenue for continued discovery of QPEs. We further note that the recent discovery of QPEs in the optical/UV transient Ansky \citep{Sanchez24,Hernandez25} indicates that the broader population of nuclear transients, including turning-on AGN and/or ``non-standard'' TDEs, also warrants further exploration. Several more QPEs are also consistent with TDEs via indirect evidence such as long-term X-ray declines \citep{Miniutti23a,Chakraborty21,Arcodia24a}, and viscously expanding accretion disks \citep{Guolo25,Wevers25} even if they are not spectroscopically confirmed. However, we limit our focus in this Letter to bona-fide optical TDEs, as their population and intrinsic rate is relatively better-understood \citep{Yao23} and provides a convenient foothold for discussions of QPE rates. Our analysis does not address all TDE classes, and only aspires toward ``completion'' in spectroscopically-confirmed optical TDEs.

With the three optical TDE/X-ray QPE sources thus far, we can make a preliminary estimate of the efficacy of targeted follow-up programs. Doing so requires a precise description of the search strategy and detection criteria, while carefully accounting for the heterogeneous information content of late-time observations across the entire population by quantifying the likelihood (in a Bayesian sense) that QPEs exist even if they are not detected. We first outline the necessary steps, then apply the method to real X-ray archival data of optically selected TDEs within $z<0.3$, roughly the horizon out to which QPEs of $L_{X,\rm peak}\lesssim 10^{44}$ erg s$^{-1}$, are detectable by current X-ray facilities. Our analysis is predicated on following assumptions:
\begin{enumerate}
    \item The fraction of optical TDEs resulting in QPEs, which we refer to as $q$, is assumed to be constant across the population, and the unique properties of each TDE and its host galaxy are ignored. While this need not necessarily be the case (see Section~\ref{subsec:ECLEs}), the goal of our analysis is to estimate the population-level average.
    \item A ``QPE detection'' is considered an observation which measures a one-off high-amplitude rapid flare of $\geq 5\times$ higher than the median flux, prompting high-cadence follow-up to confirm the recurrent nature of the bursts. While detecting a single flare does not confirm QPEs, as detection of the recurrence is also necessary, $\geq 5\times$ flares on hour-/day-timescales are not typically seen in non-jetted TDEs, and prompt follow-up of all three spectroscopically confirmed TDEs meeting the above variability criterion has gone on to reveal QPEs. Making this assumption allows us to estimate the chance detection likelihood for a given monitoring cadence.
    \item Despite not having a firmly detected recurrence time, AT2019vcb is included as a QPE, given the combination of its spectral evolution and burst profiles as detailed in \cite{Quintin23} and \cite{Bykov25} making this the simplest explanation.
\end{enumerate}

Consider an observing campaign with $N_{\rm obs}$ independent observations of summed total exposure duration $T_{\rm obs}$, which aims to monitor the X-ray variability in a \textit{single} galaxy that experienced a tidal disruption within a fixed number of years after the TDE peak. We let the binary outcomes $\mathcal{Q}$ and $\mathcal{D}$ denote whether QPEs are \textit{present} and \textit{detected}, respectively, in that galaxy:
\begin{align}
    \mathcal{Q} &\sim \begin{cases}
        0 \quad\rm if\;QPE\;phenomenon\;off \\
        1 \quad\rm if\;QPE\;phenomenon\;on \\
    \end{cases} \\
    \mathcal{D} &\sim \begin{cases}
        0 \quad\rm if\;QPE\;peak\;not\;detected \\
        1 \quad\rm if\;QPE\;peak\;detected \\
    \end{cases}
\end{align}
Our goal is to estimate $q\equiv P(\mathcal{Q}=1)$, given the data, within five years of disruption. Throughout this section $P(z=Z)$ denotes the probability that the variable $z$ takes value $Z$, while $p(z=Z|y=Y)$ denotes the probability that the variable $z$ takes value $Z$ given $y=Y$.  The five-year baseline was chosen to roughly match the beginning of ZTF, which now discovers the majority of optical TDEs; as future optical capabilities and observational baselines increase, we will extend this analysis beyond five years.

If $\mathcal{Q}=0$, then $\mathcal{D}=0$ immediately. However, if $\mathcal{Q}=1$, the probability of a detection is not 1, because the eruptions themselves comprise only a small fraction of the QPE peak-to-peak recurrence time ($T_{\rm rec}$). A detection requires either that a QPE peak occurs during the observing window, or the observation itself began during a peak. The former can be modeled via a Poisson point process, given the stochastic nature of QPE recurrence time which nevertheless has an average underlying $T_{\rm rec}$, and the aperiodicity of X-ray observations, which occur across multiple visits and are influenced by a complicated blend of factors including seasonal target visibility, telescope orbital period and visibility constraints, and the observer-requested exposure strategy. As the ``observer-QPE joint system'' is inherently stochastic, the Poisson description predicts that the expected number of QPEs within the observing window is $T_{\rm obs}/T_{\rm rec}$, thus the void probability (of no QPE peaks) is $\exp(-T_{\rm obs}/T_{\rm rec})$. The detection probability is enhanced by the chance of an observation \textit{beginning} above the detection limit, which itself is a function of $N_{\rm obs}$ and the QPE ``duty cycle'' $d$ (the fraction of $T_{\rm rec}$ spent above threshold). Combining the two probabilities, we can express the conditional odds of observing QPEs as: 
\begin{equation}
    \begin{cases}
        P(\mathcal{D}=0|\mathcal{Q}=0) = 1 \\
        P(\mathcal{D}=1|\mathcal{Q}=0) = 0 \\
        P(\mathcal{D}=0|\mathcal{Q}=1, \theta) = (1-d)^{N_{\rm obs}}e^{-\frac{T_{\rm obs}}{T_{\rm rec}}} \\
        P(\mathcal{D}=1|\mathcal{Q}=1, \theta) = 1 - (1-d)^{N_{\rm obs}}e^{-\frac{T_{\rm obs}}{T_{\rm rec}}}
    \end{cases}
\end{equation}
where $\theta\equiv (T_{\rm obs}, N_{\rm obs}, q, d, T_{\rm rec})$ describe the free parameters of the model. The total void probability of no detections, whether or not QPEs are present, is therefore:
\begin{align}
    P(\mathcal{D}=0|\theta) &= P(\mathcal{D}=0|\mathcal{Q}=0,T_{\rm obs})P(\mathcal{Q}=0) \notag \\
    &+ P(\mathcal{D}=0|\mathcal{Q}=1,T_{\rm obs})P(\mathcal{Q}=1) \notag \\
    &= (1-q) + q(1-d)^{N_{\rm obs}}e^{-\frac{T_{\rm obs}}{T_{\rm rec}}}
    \label{eq:p_d_0}
\end{align}
The converse occurs with a probability:
\begin{equation}
    P(\mathcal{D}=1|\theta) = q\big[1 - (1-d)^{N_{\rm obs}}e^{-\frac{T_{\rm obs}}{T_{\rm rec}}}\big]
    \label{eq:p_d_1}
\end{equation}
Via Bayes' theorem, the posterior probability that QPEs are present even if they are not detected, is:
\begin{align}
    P(\mathcal{Q}=1|\mathcal{D}=0, \theta) &= \frac{P(\mathcal{D}=0|\mathcal{Q}=1, N_{\rm obs})P(\mathcal{Q}=1)}{P(\mathcal{D}=0|N_{\rm obs})} \notag \\
    &= \frac{q(1-d)^{N_{\rm obs}}e^{-\frac{T_{\rm obs}}{T_{\rm rec}}}}{(1-q) + q(1-d)^{N_{\rm obs}}e^{-\frac{T_{\rm obs}}{T_{\rm rec}}}}
\end{align}
For reference, if we assume a prior QPE/TDE fraction $q=0.1$, an average peak-to-peak recurrence time $T_{\rm rec}=10$ ks, and a duty cycle (the fraction of the peak-to-peak time spent above-threshold) $d=0.1$, the posterior probability that QPEs are actually present despite a non-detection shrinks from $0.09$ with $1\times 100$ second observation, to 0.082 for $2\times 100$ sec, to 0.037 for $10\times 100$ sec.

For a single TDE, the combined likelihood function, which is given by Eq.~\ref{eq:p_d_0} if $\mathcal{D}=0$ and Eq.~\ref{eq:p_d_1} if $\mathcal{D}=1$, is:
\begin{align}
    \mathcal{L}(\mathcal{D} | \theta) &= \Big\{q\big[1 - (1-d)^{N_{\rm obs}}e^{-\frac{T_{\rm obs}}{T_{\rm rec}}}\big]\Big\}^{\mathcal{D}} \notag \\ 
    &\times \Big\{(1-q) + q(1-d)^{N_{\rm obs}}e^{-\frac{T_{\rm obs}}{T_{\rm rec}}}\Big\}^{1-\mathcal{D}}
\end{align}
To validate our analytical arguments laid out above, we performed extensive numerical simulations, which we present in Appendix~\ref{subsec:numerical_sim}.

We now turn our attention from observations of a single TDE, toward generalizing to a population of $N_{\rm TDE}$ separate TDEs. By our assumption (1) that the QPE rate is constant among all TDEs, each source can be considered an independent sampling of the same intrinsic distribution. We can therefore construct the joint likelihood function of observing QPEs of an assumed $d$ and $T_{\rm rec}$ by simply multiplying the observation outcomes of each TDE (or summing when taking the logarithm):
\begin{align}
    \log\mathcal{L}(\{\mathcal{D}_i\}|&\{N_{\mathrm{obs},i}\},\{T_{\mathrm{obs},i}\},q,T_{\rm rec},d) = \notag \\ 
    &\sum_i^{\mathcal{D}_i=1} \log\Big\{q\Big[1-(1-d)^{N_{\mathrm{obs},i}}e^{-\frac{T_{\mathrm{obs},i}}{T_{\rm rec}}}\Big]\Big\} \notag  \\
    + &\sum_i^{\mathcal{D}_i=0} \log\Big[(1-q) + q(1-d)^{N_{\mathrm{obs},i}}e^{-\frac{T_{\mathrm{obs},i}}{T_{\rm rec}}}\Big]
    \label{eq:likelihood}
\end{align}
where $i=0,1,\dots,N_{\rm TDE}$ indexes over the sources, which each have their own $N_{\mathrm{obs},i}$ and $T_{\mathrm{obs},i}$.

The total time spent observing a source, $T_{\rm obs}$, is clearly a central factor in constraining the presence of QPEs in the case of no detections. However, not all observation time can be counted equally: for example, observations with \textit{XMM-Newton} and \textit{Chandra} will probe fainter $L_{\rm peak}$, thus higher $z$, for a given source compared to \textit{Swift} or \textit{NICER}. Moreover, QPEs have been observed in at least one source to exhibit an upper threshold of the quiescence Eddington ratio \citep{Miniutti23b}, suggesting they are only observable or present for sufficiently low $\dot{M}$; this results in an effective time delay between the first detection of a TDE, and the range of observations which may even be sensitive to detecting QPEs. Thus, $T_{\rm obs}$ cannot simply be set as the total X-ray exposure on a given source, and must exclude:
\begin{enumerate}
    \item Observations which are not sensitive to faint enough fluxes to probe QPEs of a given $L_{\rm peak}$, as a result of the instrument flux limit and the source redshift.
    \item Observations which occur too soon after the TDE peak to constrain the emergence of QPEs, for some assumed time-delay between disruption and QPE emergence $\tau_{\rm delay}$.
    \item Observations which occur too long after the emergence of QPEs, for some assumed QPE lifetime $\tau_{\rm life}$.
\end{enumerate}
Thus, the free parameters $L_{\rm peak}$, $\tau_{\rm delay}$, and $\tau_{\rm life}$ modify the ``effective'' $N_{\mathrm{obs},i}$ and $T_{\mathrm{obs},i}$ of each source. Therefore, the likelihood function of Eq.~\ref{eq:likelihood} becomes a nested sum over each source \textit{and each instrument}, with only observations that probed deep enough fluxes in the correct time interval contributing.

We draw from a wide parent sample of 100 TDEs discovered by optical all-sky surveys by including every source with the ``TDE'' Object Type label and reported within the last 20 years on the Transient Name Server search tool\footnote{\href{https://www.wis-tns.org/search}{https://www.wis-tns.org/search}}. The resulting sample includes discoveries from Pan-STARRS \citep{Kaiser02}, PTF/iPTF \citep{Law09}, Gaia \citep{Gaia16}, ASAS-SN \citep{Shappee14}, ATLAS \citep{Tonry18}, and ZTF \citep{Bellm19}. We then used the \texttt{astroquery.heasarc} module \citep{Ginsburg19} to query the Goddard Space Flight Center (GSFC) High Energy Astrophysics Science Archive Research Center (HEASARC) catalog for the reported coordinates of each TDE. We saved the date and exposure duration of all observations with the \textit{XMM-Newton}, \textit{Chandra} ACIS, \textit{Swift} XRT, and \textit{NICER} telescopes, keeping only \textit{Swift} XRT observations performed in Photon Counting (PC) mode. The result was a list of every X-ray observation of every optically discovered TDE of the past 20 years, providing a near-complete seed catalog to estimate the historical probability of detecting QPEs within each source of the population.

\begin{figure}
    \centering
    \includegraphics[width=\linewidth]{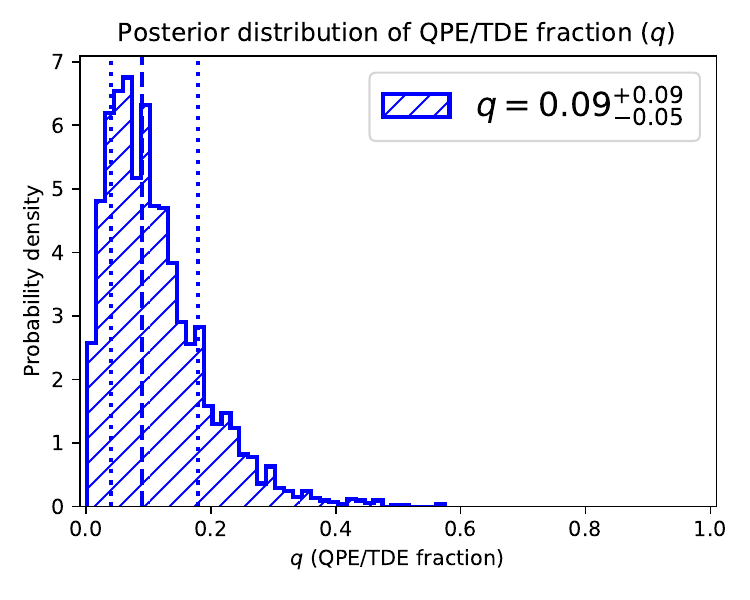}
    \caption{Posterior probability of the fraction of optical TDEs resulting in QPEs ($q$), marginalized over $T_{\rm rec}$, $\log L_{\rm peak}$, $\tau_{\rm delay}$, and $\tau_{\rm life}$. In the legend, we report the maximum a posteriori estimate and 68\% highest density credible intervals. Full corner plots can be found in Appendix~\ref{subsec:corner_plots}.}
    \label{fig:rate}
\end{figure}

Equipped with the optical TDE catalog thus described, as well as the likelihood function of Eq.~\ref{eq:likelihood}, we derived posterior probability distributions and the Bayesian evidence with the nested sampling Monte Carlo algorithm MLFriends \citep{Buchner14,Buchner21} using the UltraNest\footnote{\url{https://johannesbuchner.github.io/UltraNest/}} Python package \citep{Buchner21}. In Fig.~\ref{fig:rate}, we show the posterior probability distributions for $q$, the QPE/TDE fraction, marginalized over the uncertain distributions of $T_{\rm rec}$, $\log L_{\rm peak}$, $\tau_{\rm delay}$, and $\tau_{\rm life}$. For our sampling, we adopted a uniform prior for $q\sim\mathcal{U}(0,1)$; a uniform prior for $T_{\rm rec}\sim\mathcal{U}(0.1,5\;\rm days)$ based on the sample of known QPEs; a power-law prior $P(\log (L_{\rm peak})) \propto -1.83\log L_{\rm peak}$ with a lower/upper cut-offs at $10^{41.7}/10^{43.3}$ erg s$^{-1}$ based on the QPE luminosity function derived in \cite{Arcodia24b}; a uniform prior for $\tau_{\rm delay}\sim \mathcal{U}(0.1, 5\;\rm yrs)$; and a uniform prior for $\tau_{\rm life}\sim \mathcal{U}(1, 20\;\rm yrs)$. We assumed a conservative duty cycle of $d=0.1$, kept constant based on the relatively weak power-law scaling of the empirical $T_{\rm dur}\propto T_{\rm rec}^{4/3}$ relationship in the known QPEs (Fig.~\ref{fig:tdur_trecur}). As discussed above, for each choice of $\log L_{\rm peak}$, $\tau_{\rm delay}$, and $\tau_{\rm life}$, we censored any X-ray observations without sufficient flux sensitivity or not during the appropriate QPE detectability window. We used nominal sensitivity thresholds of $5\times 10^{-14}$ erg cm$^{-2}$ s$^{-1}$ of \textit{XMM-Newton} and \textit{Chandra}, and $5\times 10^{-13}$ erg cm$^{-2}$ s$^{-1}$ for \textit{NICER} and \textit{Swift} to ensure sensitivity to flares of $\geq 5\times$ quiescence. Marginalized over these uncertainties, we estimate that the QPE/TDE fraction is $q=9^{+9}_{-5}$\% (where the reported values are the maximum a posteriori estimate and the 68\% highest density credible interval). We show the corner plot from our sampling run, depicting the covariance of $q$ with other free parameters, in Appendix~\ref{subsec:corner_plots}.

\section{Discussion} \label{sec:discussion}
\subsection{Comparisons to QPE volumetric rates, theoretical expectations, and implications for the EMRI rate} \label{subsec:volumetric}
\cite{Arcodia24b} reported an observationally derived estimate of the volumetric rate of QPEs, based on the eROSITA-discovered QPEs (eRO-QPEs 1-4) and the all-sky survey detection efficiency and scanning strategies of eRASS 1-5. Their study provides a convenient independent estimate of the QPE intrinsic distribution, as it is an all-sky, volumetric quantity rather than a value inherently tied to the distribution function of TDEs. We briefly recap their results:
\begin{itemize}
    \item The volumetric rate of galaxies hosting QPEs was estimated to be $\mathscr{R}_{\rm gal}= 0.60^{+4.73}_{-0.43}\times 10^{-6}$ Mpc$^{-3}$ above an intrinsic average $\log L_{0.5-2.0\;\rm keV}^{\rm peak} > 41.7$, or $\mathscr{R}_{\rm gal}= 0.36^{+2.87}_{-0.26}\times 10^{-4}$ gal$^{-3}$.
    \item This implies a QPE formation rate $\mathscr{R}_{\rm gal}/\tau_{\rm life} \approx 0.6\times 10^{-7} (\tau_{\rm life}/10\;\rm yr)^{-1}$ Mpc$^{-3}$ yr$^{-1}$, where $\tau_{\rm life}$ is the \textit{a priori} unknown QPE lifetime but assumed to have a characteristic value $\tau_{\rm life}\sim 10$ yr.
\end{itemize}
Given the volumetric rate of optical TDEs is estimated to be $3.1^{+0.6}_{-1.0}\times 10^{-7}$ Mpc$^{-3}$ yr$^{-1}$ \citep{Yao23}, the QPE rate appears to be $\sim 0.2\:(\tau_{\rm life}/10\;\rm yr)\times$ smaller than the optical TDE formation rate. When accounting for the fact that only $\sim 40\%$ of optical TDEs are also X-ray detected \citep{Guolo24b}, this fraction decreases to $\sim 0.08\times$ the optical TDE rate, which is strikingly consistent with our results.

One important consideration is that the eRO-QPE sources studied by \cite{Arcodia24b} were not selected via optical TDEs, but discovered blindly in an all-sky X-ray survey. In fact, it is not clear these X-ray selected QPEs are the same population as the optical TDEs, as the presence of an optical precursor flare within 5 years prior to the eRO-QPEs, and in the years following the discovery of other known QPEs (except those selected via optical TDEs), can be ruled out \citep{Arcodia21,Arcodia24a}. In general there are large uncertainties regarding the rates and relationships of TDE populations selected in different wavelengths, including seemingly distinct populations of X-ray only \citep{Sazonov21}, infrared-only \citep{Masterson24}, radio-bright \citep{Cendes24}, optical-only, and optical-plus-X-ray \citep{Guolo24b} TDEs. It is thus interesting, and not immediately obvious, that the QPE populations emerging from different channels should be consistent with a common rate.

In the collisional model, the fraction of TDEs resulting in QPEs is primarily a function of the occupation rate of EMRIs with hour-to-day orbital periods in the galactic nuclei harboring TDEs. At these orbits, the formation rate of stellar EMRIs is expected to be dominated by the Hills mechanism, with a characteristic rate $\sim 10^{-5}$ yr$^{-1}$ for a $10^6 M_\odot$ SMBH \citep{Hills88,Linial23a}. Observationally, the rate of TDEs is $\sim$1/galaxy/$10^4$ yr \citep{Yao24}. Thus, an order-of-magnitude argument is that QPEs should occur in $\sim$10\% of TDEs \citep{Linial23b}, which is also consistent with our estimates. Future measurements, with a larger sample of QPEs and TDEs, will be able to narrow the observational error bars, from which we may hope to learn about the relevant processes placing EMRIs on gravitationally bound orbits around SMBHs in galactic nuclei \citep{Rom24,Rom25}. Some non-collisional models also make suggestions for the QPE-TDE connection, such as the unique vertical structure of the compact super-Eddington disks formed in TDEs sourcing QPEs via nodal precession \citep{Middleton25}, the presence of long-term TDE-like flares being partially fed by the same compact object that fuels the QPEs \citep{Lu23,Yang25}, or the newborn disks of TDEs being subject to unique disk instabilities not present in long-lived AGN disks \citep{Pan22}. To our knowledge, those models do not make explicit predictions for the \textit{fraction} of TDEs showing QPEs, but it is a worthwhile area for follow-up study.

\subsection{Relation to ECLEs} \label{subsec:ECLEs}
A subset of TDEs/TDE candidates are known which show unusually strong coronal emission lines, which are produced by forbidden transitions of high ionization potential $\gtrsim 90$ eV. They predominantly comprise iron transitions excited by the soft X-ray emission from the high-temperature accretion disks formed in TDEs, and must exist in low-density gas ($10^7-10^{10}$ cm$^{-3}$) to allow the forbidden transitions to occur. Typically ECLEs are identified via some combination of [Fe\,{\sc vii}] $\lambda 3759$ \AA, [Fe\,{\sc vii}] $\lambda 5160$ \AA, [Fe\,{\sc vii}] $\lambda 5722$ \AA, [Fe\,{\sc vii}] $\lambda 6088$ \AA, [Fe\,{\sc x}] $\lambda 6376$ \AA, [Fe\,{\sc xi}] $\lambda 7894$ \AA, [Fe\,{\sc xiv}] $\lambda 5304$ \AA, [S\,{\sc xii}] $\lambda 7612$ \AA. They have been noted in about eleven TDEs thus far, including AT2017gge/ATLAS17jrp \citep{Onori22}, AT2018bcb/ASASSN-18jd \citep{Neustadt20}, AT2018gn/ASASSN-18ap \citep{Masterson24,Wang24}, AT2018dyk \citep{Clark25}, AT2019qiz \citep{Short23}, AT2020vdq \citep{Somalwar23b}, AT2021dms \citep{Hinkle24}, AT2021qth \citep{Yao23}, AT2021acak \citep{Li23}, AT2022fpx \citep{Koljonen24}, and AT2022upj \citep{Newsome24}.

X-ray coverage of the majority of the 11 aforementioned ECLEs is extremely heterogeneous and generally sparse, but even among the existing observations, it is quite suggestive that \textit{two of three known optical TDEs showing QPEs are also ECLEs} (AT2019qiz and AT2022upj). Recent ECLE samples selected from the SDSS and BOSS spectroscopic surveys estimate a galaxy-normalized incidence rate of order $\sim1-3 \times 10^{-6}$ galaxy$^{-1}$ yr$^{-1}$, or a volumetric rate of $1-10 \times 10^{-9}$  Mpc$^{-3}$ yr$^{-1}$; this is one to two orders of magnitude smaller than TDE rates from the literature, consistent with variable ECLEs being caused by $f=5-20\%$ of all TDEs \citep{Callow24,Callow25}. A naive calculation of randomly drawing from the overall TDE sample, and finding that two of them are ECLEs, yields a chance coincidence probability of  ${3\choose 2}f^2(1-f)= 0.7-9.6$\%; in other words, it is possible that there is some association between QPEs and ECLEs. Moreover, many TDEs show late-time development of coronal lines (e.g. they were detected $+428$ days after the initial peak in AT2019qiz; \citealt{Short23}), so we cannot rule out the presence of coronal lines in other TDEs/QPEs. Finally, it is also noteworthy that of the remaining ECLEs, one has been suggested as a repeating partial TDE based on a second optical flare $\sim 2$ years after the initial TDE detection (AT2020vdq; \citealt{Somalwar23a}). While the processes generating rpTDEs and QPEs are probably not identical, this may indicate both rates are enhanced in ECLEs.

Multiple studies suggest that extreme coronal lines appear in those TDE which occur in particularly dust- and gas-rich environments, because 1) they show a high rate of mid-infrared echoes suggesting significant densities of circumnuclear dust reprocessing the TDE emission; and 2) the star formation rates of ECLEs lie on the high end of the TDE distribution, with a mean value of 1 $M_\odot$ yr$^{-1}$ for ECLEs compared to the overall TDE average of $\sim 0.1 M_\odot$ yr$^{-1}$ \citep{Wang12,Hinkle24}. The higher star formation rates suggest an increased abundance of nuclear gas \citep{Feldmann20}, which is not particularly surprising as substantial gas density is required for coronal line formation to occur at a detectable level. 

In the EMRI collisional picture, a QPE-ECLE coincidence may suggest that dissipative interactions with ambient nuclear gas can enhance the rate of EMRIs within $\sim$100s $R_g$ of the central SMBH (akin to the ``wet EMRI'' formation channel) relative to the ``dry'' channel, where their rate is governed by gravitational wave (GW) circularization. One possible gas-induced formation scenario involves excitation of highly eccentric orbits by angular momentum relaxation (e.g., two-body scattering, resonant relaxation), followed by rapid circularization driven by dynamical friction (DF) from the surrounding gaseous environment. DF leads to orbital circularization, provided that the ambient density falls off steeply with radius, such that the energy dissipation rate, $t_{\rm E,DF}^{-1} =|\dot{E}_{\rm orb}/E_{\rm orb}|$, dominated at pericenter passage, exceeds the angular momentum evolution rate $t_{\rm J,DF}^{-1}=|\dot{J}_{\rm orb}/J_{\rm orb}|$, dominated by dynamical friction torques at apocenter. Dynamical friction would thus contribute to EMRI formation relative to the ``dry'' channel, as long as $t_{\rm E,DF} \lesssim t_{\rm E, GW}$, where $t_{\rm E,GW}$ is the GW induced orbital circularization rate \citep[e.g.,][]{Linial23a}, with an enhanced rate of orbital dissipation. The ratio of energy dissipation timescales, evaluated at radius $r$ is given by
\begin{equation}
    \frac{t_{\rm E,DF}}{t_{\rm E,GW}} \approx \frac{M_\bullet}{\rho(r) r^3} \frac{M_\bullet}{m_\star} \left( \frac{r}{R_{\rm g}} \right)^{-5/2} \,,
\end{equation}
up to order-unity prefactors, where $M_\bullet$ and $m_\star$ are the SMBH and EMRI mass, and $\rho(r)=\rho_0 (r/r_0)^{-\gamma}$ is the assumed ambient density profile. Dynamical friction dominates over GW circularization when $t_{\rm E,DF}/t_{\rm E,GW} \ll 1$, imposing a condition on the enclosed gas mass, $\approx \rho(r)r^3 \propto r^{3-\gamma}$. For realistic values of $\gamma$, $t_{\rm E,DF}/t_{\rm E,GW} \propto r^{\gamma-11/2}$ is a decreasing function of $r$, suggesting that DF circularization dominates as long as the total gas mass enclosed within some reference radius $r_0$ satisfies $\rho_0 r_0^3 \gg M_\bullet (M_\bullet/m_\star) (r_0/R_{\rm g})^{-5/2}$. Given the density scale implied by the presence of strong coronal emission lines, $\rho_0 \approx m_{\rm p} \times 10^{8.5} {\rm cm^{-3}}\approx 5\times10^{-16} \, \rm g \,cm^{-3}$, we infer a lower limit on the radial scale harboring these gas densities, of order $r_0$ is $r_0 \gtrsim 10^{16} \, \rm cm \; (M_\bullet/10^6 \, \rm M_\odot)^{9/11}$, amounting to $\gtrsim 0.5 \, \rm M_\odot$ in gas mass on this scale. We also note this length scale is $\gtrsim 100\times$ the accretion disk radial extent of $\sim 10^{14}$ cm, as it must be to significantly impact the delivery of EMRIs onto sufficiently tight orbits for QPEs. A systematic examination of the accretion disk sizes, and the inferred radial scales/implied gas densities, could be an interesting avenue to determine whether galaxies in which QPEs are observed all fulfill a similar condition.

\section{Conclusion} \label{sec:conclusion}
We reported the discovery of quasi-periodic eruptions (QPEs) in the optically discovered Tidal Disruption Event (TDE) and Extreme Coronal-Line Emitter (ECLE) AT2022upj (Fig.~\ref{fig:lc}). The QPEs exhibit soft blackbody-like spectra with characteristic temperatures of $kT \sim 80-210$ eV (Fig.~\ref{fig:spec}, Table~\ref{tab:spec}), and have a large spread in both recurrence time (0.5-3.5 days) and duration (0.3-1 days). Phase-resolved spectroscopy of individual eruptions revealed significant spectral evolution over each flare, with temperatures tending to rise rapidly during the flare onset and then drop throughout the decline (Fig.~\ref{fig:spec_evol}). The flare profiles are relatively erratic, with some showing the usual rapid rise and slower decline, but with some overlapping bursts (Fig.~\ref{fig:lc}) as noted in e.g. eRO-QPE1 \citep{Arcodia22}. Using the optical/UV plateau and late-time X-ray emission, we estimated the properties of the viscously expanding accretion disk formed in AT2022upj using the {\tt FitTeD} code of \cite{Mummery24b}. We found that the multi-band light curve is indeed consistent with the expected radial scale in the EMRI-disk collisional model, i.e. a $\sim 0.5M_\odot$ accretion disk which viscously expands to reach the orbital radius producing $\sim 3$-day QPEs around the time they were first observed (Fig.~\ref{fig:disk_model}). The independent SMBH mass estimate produced by this fit, $M_\bullet \sim 5.6^{+3.3}_{-2.5}\times10^6M_\odot$, is slightly higher than previous mass estimates from \cite{Newsome24} (which, however, are mostly drawn from scaling relations with large scatter). Despite the large in flare profiles, the duration and recurrence times align with the overall $T_{\rm dur}\propto T_{\rm rec}^{1.3}$ noted in the growing QPE population (Fig.~\ref{fig:tdur_trecur}).

With the addition of AT2022upj, three optically discovered TDEs have now shown QPEs at late times. Motivated by this, we made a probabilistic Bayesian argument to estimate the fraction of TDEs that host QPEs, accounting for heterogeneous X-ray follow-up data across a large sample of optical TDEs. Marginalizing over possible distributions for recurrence time, QPE peak luminosity, the delay between TDE peak and QPE onset, and QPE lifetime, we infer $q = 9^{+9}_{-5}$\% of TDEs produce QPEs within five years of disruption (Fig.~\ref{fig:rate}). Our preliminary estimate agrees with an independent measurement of the QPE volumetric rate, which suggests based on the the volumetric rates of X-ray QPEs \citep{Arcodia24b} and X-ray+optically bright TDEs \citep{Yao23,Guolo24b} that $\sim 8\%$ of optical TDEs may show QPEs (Section~\ref{subsec:volumetric}).

Moreover, AT2022upj is the second QPE discovered in an ECLE (after AT2019qiz). Given that ECLE and TDE volumetric rates suggest only $\sim 5-20$\% of TDEs show coronal lines \citep{Callow25}, the naive coincidence probability of two of three optical TDEs showing QPEs being randomly associated with ECLEs is $0.7-9.6$\%. These low odds suggest that there may be a link between the formation of QPEs and ECLEs, though the small-number statistics at this stage limits this claim to a speculation to be tested. Multiple studies suggests that ECLEs are TDEs in gas-rich galaxy nuclei \citep{Wang12,Hinkle24}, based on their enhanced star formation rates and prevalence of dust reprocessing echoes; in this scenario, the dissipative effects from the enhanced nuclear gas densities may account for a possible higher rate of EMRIs of hour-to-day orbits in the central $\sim$hundreds $R_g$. The opposite causal direction is also possible, and the sustained high-energy of emission may be responsible for exciting the coronal line emission seen in some ECLEs; further time-resolved multiwavelength studies of both QPEs and ECLEs will elucidate this relationship.

Our findings contribute to the growing body of evidence linking QPEs to newly formed TDE accretion flows, possibly when a stellar-mass object (the EMRI companion) intersects a nascent accretion disk in repeated collisions that generate the soft X-ray bursts. While our error bars on the QPE-TDE rate are still large given the small-number statistics, future work will benefit from larger samples of TDEs discovered by next-generation synoptic surveys \citep{LSST19} monitored over increasing baselines. Upcoming and proposed X-ray observatories---such as the Advanced X-ray Imaging Satellite (AXIS; \citealt{Reynolds24}), which will have $\sim 100\times$ greater sensitivity than \textit{Swift} and \textit{NICER}, as well as faster slew and longer monitoring capabilities than \textit{XMM-Newton}---will be crucial in extending our reach to a larger sample of QPEs at higher \textit{z} and longer timescales. Systematic searches will yield valuable insights into the nature of these transient accretion events and help constrain astrophysical processes in galactic nuclei, including stellar dynamics and capture, accretion disk formation evolutions, and the rates and properties of extreme mass-ratio inspirals.

\section*{Acknowledgements}
We thank the \textit{NICER} scheduling and operations team for prompt, high-cadence monitoring of AT2022upj, which enabled this discovery. We thank the anonymous referee for comments which greatly improved the quality of the manuscript.
MG is funded by Spanish MICIU/AEI/10.13039/501100011033 grant PID2019-107061GB-C61.
GM thanks the Spanish MICIU/AEI/10.13039/501100011033 and ERDF/EU grant PID2023-147338NB-C21 for support. IL acknowledges support from a Rothschild Fellowship and The Gruber Foundation, as well as Simons Investigator grant 827103. 
LHG acknowledges funding from ANID programs:  FONDECYT Iniciación 11241477, and Millennium Science Initiative Programs NCN$2023\_002$ and AIM23-0001.

\bibliography{refs}{}
\bibliographystyle{aasjournal}

\appendix
\counterwithin{figure}{section}
\counterwithin{table}{section}

\section{Observations and Data Reduction} \label{sec:methods}
\subsection{NICER} \label{subsec:nicer} 
Based on an ongoing program to obtain late-time coverage of a wide sample of tidal disruption events, we requested Target-of-Opportunity observations on December 12th, 2024. The \textit{NICER} X-ray Timing Instrument \citep{Gendreau16} aboard the International Space Station subsequently observed AT2022upj for a total of 38.3~ks across 23 Target of Opportunity (ToO) observations (ObsIDs 7203550101-7203550123) from Dec. 12, 2024-Jan. 8, 2025. We followed the time-resolved spectroscopy approach for reliable estimation of source light curves outlined in Section 2.1 of \cite{Chakraborty24}.

Spectral fitting and background estimation was performed with the \texttt{SCORPEON}\footnote{\href{https://heasarc.gsfc.nasa.gov/docs/nicer/analysis_threads/scorpeon-overview/}{https://heasarc.gsfc.nasa.gov/docs/nicer/\\analysis\_threads/scorpeon-overview}} model over a broadband energy range (0.25--10 keV) for data taken in orbit night, and a slightly restricted range (0.38--10 keV) during orbit day. \texttt{SCORPEON} is a semi-empirical, physically motivated background model which explicitly includes components for the cosmic X-ray background as well as non X-ray noise events (e.g. precipitating electrons and cosmic rays) and can be fit along with the source to allow joint estimation of uncertainties. We grouped our spectra with the optimal binning scheme of \cite{Kaastra16}, i.e. \texttt{grouptype=optmin} with \texttt{groupscale=10} in the \texttt{ftgrouppha} command, and performed all spectral fitting with the Cash statistic \citep{Cash1979}. The light curve thus generated is presented in Fig.~\ref{fig:lc}.

\subsection{ZTF and Swift} \label{subsec:ztf_swift}
The Zwicky Transient Facility (ZTF; \citealt{Bellm19}) has been systematically conducting a wide-field survey of the northern sky from Palomar Observatory since 2018. 
We retrieved data from the ZTF Forced Photometry Service in the g- and r-bands \citep{masci2023}. The measurements obtained from this service are obtained from the reference-subtracted images, therefore subtracting the host galaxy emission and isolating the variable component.
The light curves were constructed with the criteria explained in \cite{masci2023} and in \cite{lore2023}, including quality filtering and rejecting bad-data quality flags. We did not apply color correction.
The difference PSF magnitudes were converted into apparent magnitudes following \cite{forster2021}. 
After obtaining the ZTF light curve of AT\,2022upj in AB magnitude, we converted to flux density using r-band/g-band pivot wavelengths of 6417.10/4783.50 \AA, respectively. We converted from flux density to flux by integrating over a constant filter profile from 5700-7200 \AA\ for r-band and 4200-5400 \AA\ for g-band.

We obtained ultraviolet data in the UVW1 band from the UV/optical Telescopes (UVOT) onboard the \textit{Neil Gehrels Swift Observatory} \citep{Gehrels04} via the Goddard Space Flight Center (GSFC) High Energy Astrophysics Science Archive Research Center (HEASARC). We performed aperture photometry using the \texttt{uvotmaghist} FTOOL with an 8'' source region and 30'' background region. We converted the flux densities produced by \texttt{uvotmaghist} into fluxes by integrating over a constant wavelength coverage of 2200-3200 \AA, the approximate response of the \textit{Swift} UVW1 filter.

We also obtained data from the \textit{Swift} X-ray telescope (XRT) from the online interface (\href{https://www.swift.ac.uk/LSXPS}{https://www.swift.ac.uk/LSXPS}) to the Living \textit{Swift} XRT Point Source (LSXPS) Catalogue \citep{Evans23a}. LSXPS is an automatically updated repository of all \textit{Swift} XRT observations of $>100$ second duration in Photon Counting (PC) mode. LSXPS reports X-ray fluxes in counts per second (cps); we converted to cgs units by using a conversion factor of $1\;\rm cps=2.2\times 10^{-11}$ erg cm$^{-2}$ s$^{-1}$, which was computed for a 100 eV blackbody spectrum and Galactic neutral absorption with $N_{\rm H}=2.13\times 10^{20}$ cm$^{-2}$ using WebPIMMS.  We show the long-term X-ray/optical light curve of AT2022upj using these data in Fig.~\ref{fig:lc}.

We constructed a stacked X-ray spectrum using observations from the delayed X-ray rise between MJDs 60200-60600 using the online \textit{Swift} XRT product building tool (\href{https://www.swift.ac.uk/user_objects/}{https://www.swift.ac.uk/user\_objects/}) developed and maintained by the UK Swift Science Data Centre at the University of Leicester \citep{Evans09}. This resulted in a total 72.8~ks exposure across 22 OBSIDs (00015403019-00015403041). The spectrum was grouped with a minimum of one count per bin, and subsequently fit in \texttt{XSPEC} using the Cash statistic.

\section{Numerical simulations of detection likelihood} \label{subsec:numerical_sim}
\begin{figure}
    \centering
    \includegraphics[width=\linewidth]{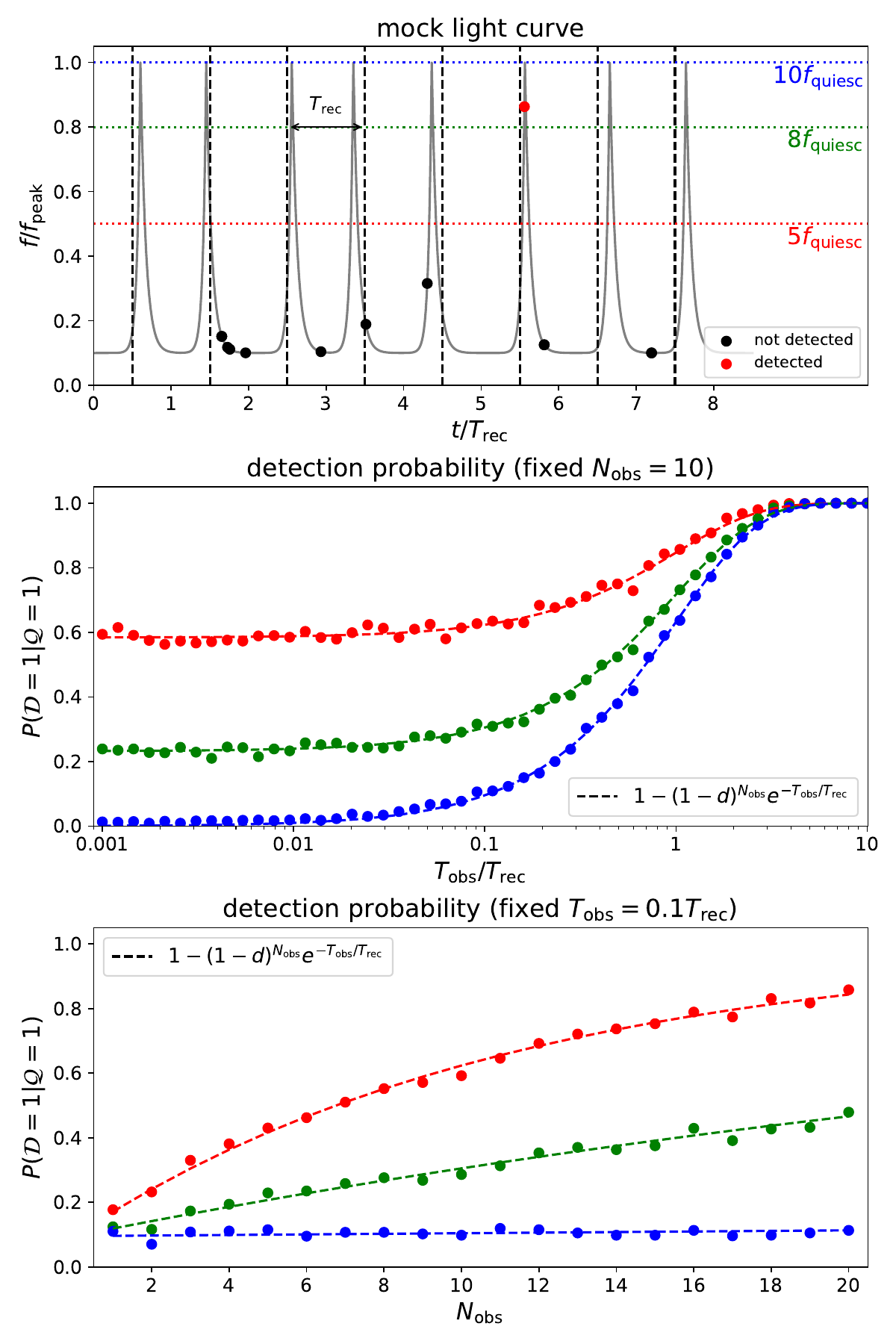}
    \caption{Numerical simulations of the QPE blind detection probability. \textbf{Top panel:} one realization of a mock light curve showing 10 randomly distributed observations each of $0.001T_{\rm rec}$ duration, of which one ``detected'' a QPE by measuring a point above the $5\times/8\times/10\times f_{\rm quiesc}$ threshold in red/green/blue, respectively. \textbf{Middle panel:} Simulated detection probabilities, with $3\times 50\times 1000$ realizations for various choices of detection threshold and total exposure time. We fix $N_{\rm obs}=10$. \textbf{Bottom panel:} same as above, for for fixed $T_{\rm obs}=0.1T_{\rm rec}$ and varying $N_{\rm obs}$.}
    \label{fig:detection_sim}
\end{figure}

To verify the probabilistic arguments laid out in Sec~\ref{sec:qpe_tde_rate}, we performed a grid of numerical simulations injecting exponential rise-and-decline profiles into a constant baseline, with a uniform $20\%$ random scatter upon a fixed recurrence time $T_{\rm rec}$, modeling the light curve behavior of a mock QPE. We then ran a trial with $N_{\rm obs}=10$ uniform random ``observations'' each with a fixed duration $T_{\mathrm{obs},i}=0.001T_{\rm rec}$, and flagged as a ``detection'' any point which coincided with a time period where the model light curve flux was $\geq 5\times$ the baseline value. An example mock light curve, and set of observations with one detection, is shown in the top panel of Fig.~\ref{fig:detection_sim}. We ran this experiment for 1000 iterations, saving the fraction which resulted in at least one detection as ``successfully detecting QPEs'' ($\mathcal{D}=1$). We repeated the same experiment for a grid of 50 observation durations log$_{\rm 10}$-uniformly distributed between $[0.001T_{\rm rec},T_{\rm rec}]$, and for three different choices of the minimum flux threshold to claim a detection ($5\times$ in red/$8\times$ in green/$10\times$ in blue). As expected, for higher thresholds---thus, smaller duty cycle ($d$) in which to catch an eruption---the detection probability approaches the Poisson distribution, as $\lim_{d\rightarrow 0} (1-d)^{N_{\rm obs}} = 1$ (middle panel of Fig.~\ref{fig:detection_sim}).

We also ran the experiment for $3\times50\times 1000$ iterations of varying $N_{\rm obs}$, while keeping the total summed exposure across all $N_{\mathrm{obs},i}$ fixed to $0.1T_{\rm rec}$. Again the numerical experiments agree precisely with our analytical probability function, verifying the correct dependence on $d$, $N_{\rm obs}$, and $T_{\rm obs}$ (bottom panel of Fig.~\ref{fig:detection_sim}). This latter grid of experiments numerically validates the (perhaps obvious) statement that, for a fixed total invested exposure time, the optimal strategy to detect QPEs is to spread the exposure across as many snapshots as feasible. The power-law scaling of $(1-d)^{N_{\rm obs}}$ generously rewards such an observing strategy.

\section{Supplementary figures and tables} \label{subsec:corner_plots}

\begin{figure}
    \centering
    \includegraphics[width=\linewidth]{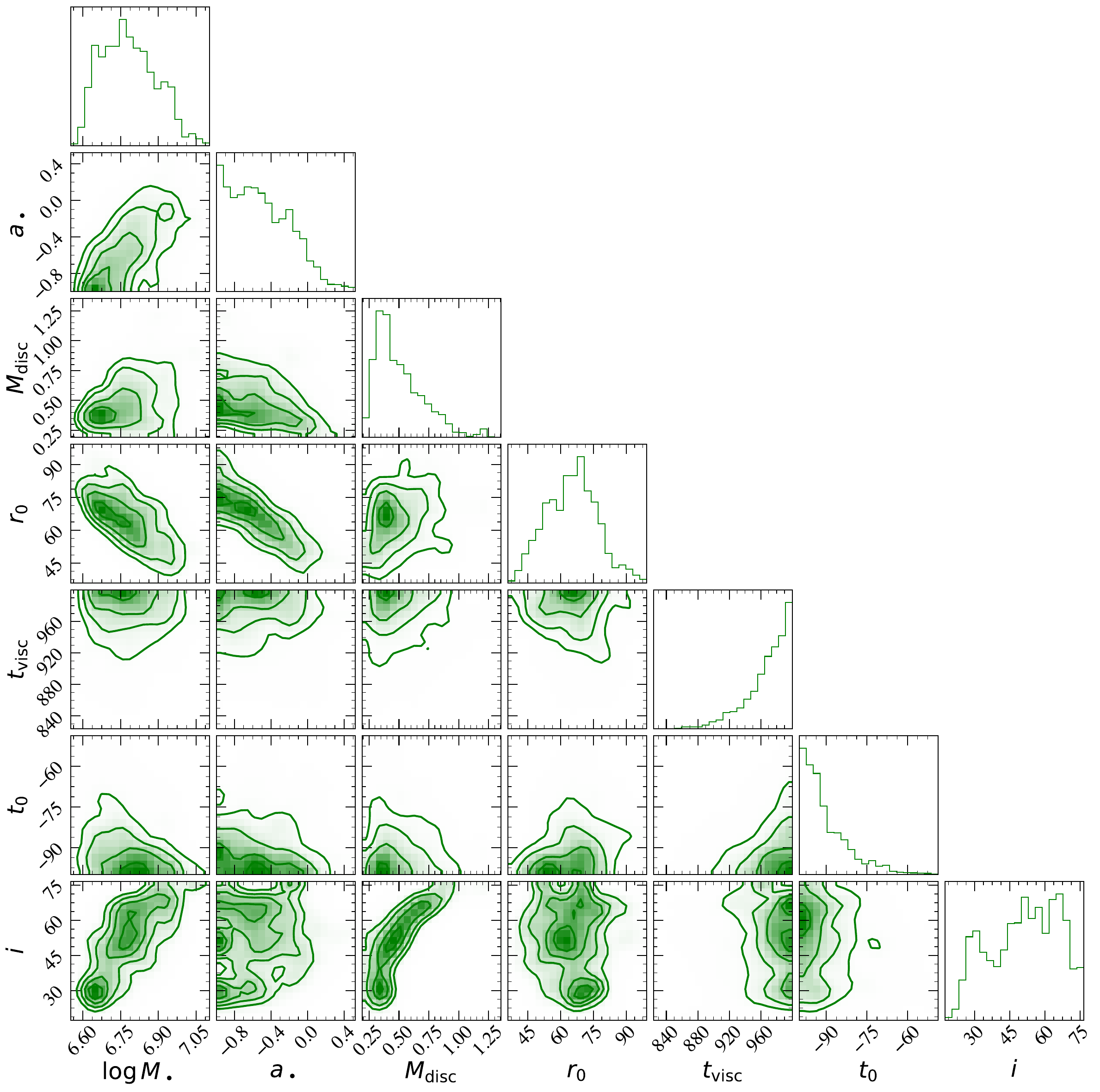}
    \caption{Corner plot for the accretion disk model fit to the multi-band light curve of AT2022upj. The resulting surface density profile and disk luminosity are shown in Fig.~\ref{fig:disk_model}.}
    \label{fig:disk_model_corner}
\end{figure}

In Fig.~\ref{fig:disk_model_corner} we show a corner plot of the relativistic thin disk model developed in \cite{Mummery20,Mummery24a,Mummery24b} fit to the multi-band data of AT2022upj. The model free parameters include central black hole mass and spin ($\log M_\bullet$ and $a_\bullet$), disk mass solar units ($M_{\rm disc}$), disk initial radius $r_0$ computed at a time $t_0$, the viscous timescale ($t_{\rm visc}$) which sets the evolution rate of the disk density, and observer inclination angle ($i$) from the disk rotation axis. Detailed descriptions of the dependence of the output model on each of these parameters can be found in \cite{Mummery24b}.

\begin{figure}
    \centering
    \includegraphics[width=\linewidth]{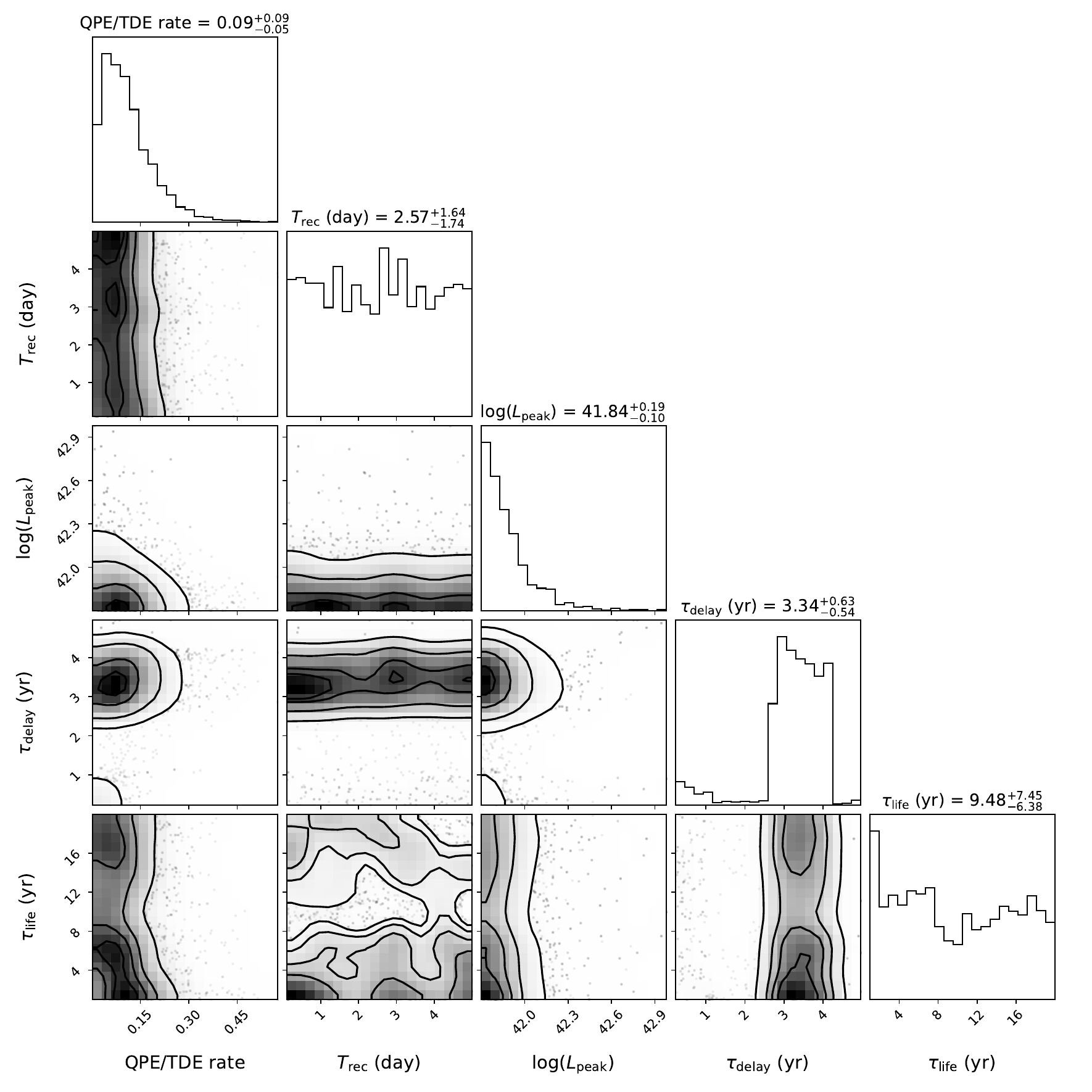}
    \caption{Corner plot for sampling run of QPE/TDE fraction.}
    \label{fig:corner_tau}
\end{figure}

In Fig.~\ref{fig:corner_tau} we show the corner plot for our sampling run to estimate the QPE/TDE fraction. The 2D histograms are smoothed by a $1.5\sigma$ Gaussian kernel for visual clarity. We find that a fraction $0.09^{+0.09}_{-0.05}$ of optical TDEs result in QPEs. There is a slight negative covariance between the inferred QPE lifetime and $\log L_{\rm peak}$: this occurs because there are effectively fewer observations capable of constraining lower-luminosity QPEs, due to the sensitivity limits of X-ray telescopes and the redshift distribution of the parent TDE sample.

\end{document}